\documentclass[aps,pre,twocolumn,showpacs,groupedaddress]{revtex4}
\usepackage{graphicx}  
\usepackage{dcolumn}   
\usepackage{bm}        
\usepackage{amssymb,amsmath}   
\usepackage{amsthm}
\usepackage{subfigure}

\newcommand{\nop}[1]{}

\begin{document}

\title{Robustness of Complex Networks against Attacks Guided by Damage}
\author{Hui Wang$^1$}
\author{Jinyuan Huang$^2$}
\author{Xiaomin Xu$^3$}
\author{Yanghua Xiao$^3$\footnote{Correspondence Author: shawyh@fudan.edu.cn}}
\author{Wei Wang$^3$}

\affiliation{$^1$ School of Economics and Management, Shanghai
University of Electronic Power, Shanghai 200090, PR China}

\affiliation{$^2$ School of Management, University of Shanghai for
Science of Technology, Shanghai 200090, PR China}

\affiliation{$^3$ School of Computer Science, Fudan University,
Shanghai 200433, PR China}

\date{\today }

\begin{abstract}
Extensive researches have been dedicated to investigating the
performance of real networks and synthetic networks against random
failures or intentional \emph{attack guided by degree} (degree
attack). Degree is one of straightforward measures to characterize
the vitality of a vertex in maintaining the integrity of the network
but not the only one. Damage, the decrease of the largest component
size that was caused by the removal of a vertex, intuitively is a
more destructive guide for intentional attack on networks since the
network functionality is usually measured by the largest component
size. However, it is surprising to find that little is known about
behaviors of real networks or synthetic networks against intentional
\emph{attack guided by damage} (damage attack), in which adversaries
always choose the vertex with the largest damage to attack.

In this article, we dedicate our efforts to understanding damage
attack and behaviors of real networks as well as synthetic networks
against this attack. To this end, existing attacking models,
statistical properties of damage in complex networks are first
revisited. Then, we present the empirical analysis results about
behaviors of complex networks against damage attack with the
comparisons to degree attack. It is surprising to find a cross-point
for diverse networks before which damage attack is more destructive
than degree attack. Further investigation shows that the existence
of cross-point can be attributed to the fact that: degree attack
tends produce networks with more heterogenous damage distribution
than damage attack. Results in this article strongly suggest that
damage attack is one of most destructive attacks and deserves our
research efforts. Our understandings about damage attack may also
shed light on efficient solutions to protect real networks against
damage attack.

\end{abstract}

\pacs{89.75.Fb, 89.75.Hc, 89.65.-s}

\maketitle

\section{Introduction}
In the last decade, great efforts have been dedicated to the
research on the resilience of real-world networks or synthetic
networks against random failures or intentional
attacks~\cite{cohen00,cohen01,paul,tani05,tani06,holme,albert,karrer,moreira,paolo,onion,cata,comm,incomp}.
Random failure can be considered as a special case of intentional
attack when no information of vertex importance in the sense of
maintaining the integrity of the network is available to an
adversary. If certain structural information of the network is
available, a rational adversary generally tends to select the most
important vertex to attack so that the destructive effect can be
maximized. Usually, the attack will continue step by step until the
adversary believes that the desired destructive objective is
achieved.

Thus, from an adversary's perspective, ranking the importance of
vertex in the network is one of the fundamental steps towards
destructing the network. A real network can be precisely modeled as
a graph $G(V,E)$, where $V$ represents the entities in the network
and $E$ represents relations among these entities. A variety of
measures are available to rank the vertex importance in a graph.
Among them, \emph{degree} and \emph{betweenness}, have been widely
investigated in previous researches about network
robustness~\cite{paolo,holme}. Degree quantifies the number of
connections to a vertex. Vertices with large degrees dominate the
connections of the whole network, thus are naturally regarded as the
most important vertex in maintaining the integrity of the network.
Betweenness counts the fraction of shortest paths going through a
given vertex. Hence, vertex with high betweenness is important in
maintaining the communication functionality of the
network~\cite{rob_comm,find_comm,alg_comm}.

In general, vertex importance can be measured from different
perspectives. However, except degree and betweenness, many of others
have been rarely investigated in the research about network
robustness. One of them is \emph{damage}, which characterizes the
damage caused by the removal of a vertex and usually is quantified
as the decrease of the largest component size when the vertex and
its incident edges are removed. Intuitively, a vertex causing large
damage leads to more destructive effect on the network performance.
Damage plays a vital role in characterizing the essentiality of
components in biological networks. Vertices (proteins or enzymes)
that cause large damage are essential or important in protein
interaction networks or metabolic networks~\cite{dam-meta,dam-ppi}.

\begin{figure}
\centering {
\includegraphics[scale=0.4]{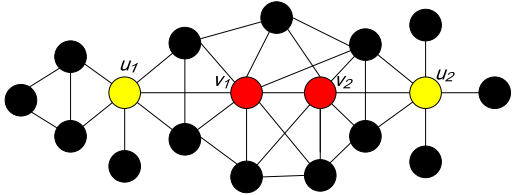}}
 \caption{(Color online) Damage attack on a hypothetical network.
 The two yellow vertex are the top two vertex of the largest damages, which are 4 and 3, respectively.
 The two red vertex are the top two vertex of the largest degrees, which are 8 and 7, respectively.}
\label{fig:example_net} 
\end{figure}

By definition, attacking the target node with the maximal damage causes
the most significant damage on the given network than any other
attack if destructive effect is measured by damage. We use a
hypothetical graph shown in Figure~\ref{fig:example_net} to
illustrate the destructive effect caused by attacking vertex of
highest damage. In this graph, vertex $u_1$ has damage value 4. The
removal of $u_1$ will isolate four vertex from the major component
of the network. However, if we remove the vertex with the largest
degree, i.e., vertex $v_1$, no vertex except itself will be isolated
from the major component. This example clearly shows that
\emph{attack guided by damage} or simply \emph{damage attack} (that
is always attacking the vertex with highest damage) yields
non-trivial destructive effect.

Despite of its destructive effect caused by damage attack, the
fundamental characteristics of damage attack and the response of
real networks as well as synthetic networks to damage attack has
been rarely studied so far. In this article, we systematically
investigate damage attack and the performance of real networks and
synthetic networks including Baraba\'si-Albert (BA)~\cite{ba}
networks and Erd\"os-R\'eny (ER)~\cite{er} networks against damage
attack. Across the study, attack guided by degree is also
investigated as a comparison.

The structure of the remainder of this article is as follows: we
first review existing attacking model aiming at unifying existing
attack models in Section II. Then, in Section III, we systematically
investigate statistical properties about damage, damage
distribution, correlation between damage and degree. In Section IV,
we present the empirical analysis results about the behaviors of
real networks and synthetic networks against damage attack with the
comparison to degree attack, finding that \emph{there exists a
cross-point before which damage attack is more destructive than
degree attack}. Such findings are further explained in Section V.
Finally, in Section VI, we briefly summarize major findings and
results that we get in this article.

\section{Attacking Models}
A variety of intentional attacking models have been implicitly
studied in previous researches. Despite of their distinctive
features, they share the same framework to describe themselves. In
general, we need to specify the ranking mechanisms of vertex
importance and performance measure of networks to clarify an
attacking model. In this section, we will first present a unified
attacking model by synthesizing existing models, and then give a
brief overview of vertex importance ranking and network
functionality measures.

\subsection{Unified attacking model}
An attacking strategy describes the way that an adversary issues
attacks on the network. In this paper, we are particularly
interested in \emph{intentional attack}, since intentional attack is
more destructive than random attack or random
failures~\cite{albert,cohen01, paul,tani05}. In an intentional
attack, an adversary preferentially attacks the vertex that he
believes is important for maintaining integrity of the network to
maximize the destructive effect. In contrast, in a random attack or
random failure, adversaries choose the attack target by purely
random guess or each vertex or edge fails with equal probability.

Among intentional attacks, we only consider \emph{vertex attack},
that is, attacking a network by removing a vertex as well as its
incident edges from the network. Some conclusions can be directly
extended to edge attack, i.e., just removing an edge. It's
reasonable to assume that the objective of an adversary to attack a
network is to maximize the destructive effect by attacking fixed
number of vertices. For this purpose, an adversary first need to
assess the \emph{'benefits'} by attacking an vertex. This problem is
equivalent to evaluate the importance of a vertex for maintaining
the integrity of the network. Hence, ranking mechanisms of vertex
importance becomes a core of devising an attacking strategy. In
general, different adversaries will rank the vertex importance from
different aspects, producing disparate destructive effects. For
example, the vertex importance can be measured by degree,
betweenness, damage or many others.

Suppose the adversaries are empowered with the ability to identify
the most important vertex, he generally will follow a
\emph{greedy} framework to maximize the destructive effect, that is
\emph{always attacking the most important vertex of the remaining
network at each step until he believes that the desired destructive
effect is achieved}. This is the framework of attack model that will be
investigated in this paper.

Note that in above attacking model, we make two strong assumptions. The first is that adversaries can always select the most important vertex. The second is that adversaries have the
ability to issue continuous attacks. Such
strongest assumptions are quite meaningful since protection solution
based on the understanding of the network robustness behavior
against the worst attack comes with safety guarantee. In many real
cases, adversaries can hardly select the most important vertex
since only local information about the network is available to
adversaries~\cite{holmeepl}. On the hand, adversaries usually have limited resources to issue continuous attack. Hence, in real cases the networks preform better than we expected under the strongest assumption.

Another key element of an attacking model is the quantification of
network performance or functionality. The characterization of the
robustness of networks is determined by performance measures. For
example, the robustness of network can be investigated by observing
the change of the largest component size ($S_{max}$) of the network
when the network is subject to continuous attacks~\cite{albert,holme07}. Fast decrease of
$S_{max}$ provides strong evidence for the vulnerability of the
network. The network performance can also be measured from other
aspects, e.g., the communication functionality, which are usually
measured by the average shortest path length or network diameter. In
general, when measured from different aspects, a network usually
exhibits diverse robustness behaviors.

Hence, generally, an attacking model can be described by a triple
$(\alpha, \tau, $F$)$, where $\alpha$ is the functionality measure,
$\tau$ captures the most important vertex as the attacking target
and $F$ is an indicator function that indicates whether the
attacking objective is achieved. Let $G_t$ be the remaining network
after $t$ steps of attack, then the attacking procedure specified by
$(\alpha, \tau, F)$ generally can be described as follows:\emph{ At
each attacking step, $\tau(G_t)$ as well as its incident edges are
removed from $G_t$. Repeat the attack until $\alpha(G_t)$ satisfies
the assertion specified by $F$.}

In most previous researches, when $\alpha(G)$ is the largest
component size of a network, $F$ is assumed to be $\alpha(G_t)=0$,
or $\alpha(G_t)\approx 0$. In cases where $F$ is clear in the
context, we usually represent an attacking model by $(\alpha,
\tau)$. Note that, in the proposed attacking model, $\tau(G_t)$ is
always calculated from $G_t$ instead of $G_0$, which means we always
recalculate the vertex importance and select one of the most
important from the network after attack instead of the original
network. The rationality lies in the fact that most measures of
vertex are sensitive to vertex removal. However, when attack is
guided by degree, removing vertex with the highest degree has
limited influence on the degree rank of remaining vertex. Hence, we
can use the degree rank of the original network to approximate that
of the network after attack.

Under the attack specified by $(\alpha, \tau)$, the robustness of
a network $G$, denoted by $r(G)_t$, can be explicitly defined as:
\begin{equation}
r(G)_t=\frac{|\alpha(G_0)-\alpha(G_t)|}{t} \label{equ:robustness}
\end{equation}, which is the ratio of network performance decrease to the number
of removed vertex. From the viewpoint of adversaries, it
characterizes the efficiency of an attack strategy.

\subsection{Measures of network functionality}
In the study of network robustness, a typical measure of network
functionality is the size of its largest connected component
(LCC)~\cite{albert}, i.e., $S_{max}$. In general, removing a vertex
from a connected network will detach some connected subgraphs or
isolated vertex from the largest component of the network. Hence,
$S_{max}$ will decrease after deleting a vertex. Then, the
vulnerability or robustness of a network can be evaluated by the
change of $S_{max}$. More larger the change of $S_{max}$ is, more
vulnerable the network is. When we investigate damage attack, we
will use $S_{max}$ as the major measure for network functionality.

Another class of network functionality measures is based on the
shortest path information of the network. Shortest paths have been
shown to be crucial for the communication of a variety of real
networks. Thus, preserving the key statistics of shortest pathes,
will be an important indicator of resilience of the network against
vertex attack. One statistic over shortest paths in a network
$G(V,E)$ is the average shortest path length, which is usually
defined as:
\begin{equation}l=\frac{1}{N(N-1)}\sum_{u,v\in V,
u\neq v}d(u,v)
\end{equation}, where $d(u,v)$ is the length of the shortest path
from vertex $u$ to $v$. In a disconnected network, $d(u,v)$ will be
infinite for two disconnected vertices $u$ and $v$. To overcome this
problem, one can instead use the average inverse geodesic length,
which is
\begin{equation}l^{-1}=\frac{1}{N(N-1)}\sum_{u,v\in V, u\neq
v}\frac{1}{d(u,v)}
\label{equ:spi}
\end{equation} When attacking a network by removing vertices,
the network will eventually break into disconnected subgraphs.
Hence, $l^{-1}$ as given in Equation~\ref{equ:spi} is also widely
used to measure network functionality. When performing continuous
attack on a network, $l^{-1}$ will decrease with the increase of the
number of removed vertex. Fast decrease of $l^{-1}$ strongly
suggests that the network is vulnerable.

\subsection{Ranking of vertex importance}
A variety of vertex importance measures have been proposed, among
them, degree and betweenness are two most widely used measures.
Intuitively, vertices with high degree, i.e. \emph{hub vertices},
contribute significantly to the interconnectedness of the whole
network, thus is important with respect to maintaining the integrity
of the network. Vertex betweenness measures the number of shortest
paths passing through a vertex. Vertices with high betweenness are
believed to be crucial for information/material transferring in many
real networks such as internet, power grid, where information or
signals travels from sources to destinations by shortest paths to
save the transfer cost. The vertex betweenness of a vertex $u$ can
be formally defined as follows:
\begin{equation}
C_B(u)=\sum_{s\neq u\neq t\in V}\frac{\sigma_{st}(u)}{\sigma_{st}}
\end{equation}, where $\sigma_{st}$ is the number of shortest paths between $s$ and $t$ and $\sigma_{st}(u)$ is the
number of shortest paths in $\sigma_{st}$ that pass through $u$ .

One important issue about vertex ranking is its sensitivity to the
vertex removal operation. Clearly, ranking by degree is less
sensitive than ranking by betweenness to vertex removals. Such a
fact implies that recalculation on destructed networks is necessary
for betweeness-based ranking. In contrast, degree ranking in the
original network is quite close to that obtained by recalculating on
destructed networks. In general, adversaries need to pay extra
recalculation cost to perform attacks guided by vertex ranking
mechanisms that are sensitive to vertex removals.

Computation cost is another critical concern when adversaries attack
a network consisting of millions of vertex. Degree can be retrieved
by constant time when the graph is built in memory. The fast
algorithm to compute betweenness for all nodes requires $O(NM)$ time
for unweighted networks~\cite{fastbw}, where $N, M$ are the vertex
number and edge number, respectively. However, degrees of all vertex
can be counted in $O(N+M)$ time. Hence, considering computation
cost, degree is preferred to betweenness by adversaries. However,
for small networks or medium-sized networks of tens of thousands of
vertices, both degree and betweenness can be obtained by
adversaries, which poses a great challenge for us to protect real
network systems.

\section{Damage of networks}

If the network functionality is measured by $S_{max}$, then to
maximize the destructive effect, a straightforward greedy approach
is to remove vertices one by one in the descending order of their
damage value. In this section, we will systematically revisit vertex
damage for real networks and synthetic networks.

\subsection{Damage of graphs}
Let $G(V,E)$ be an undirected graph (not necessarily a connected
graph). The damage of a vertex $v\in V$, $D(v)$ is defined as
$S_{max}-S'_{max}$, where $S_{max}$ and $S'_{max}$ are the largest
connected component size before removing $v$ and after removing $v$,
respectively. For a vertex $v$ in a connected graph, its damage
$D(v)$ lies in the range $[1, N-1]$\footnote{Note that when the
graph is not connected, the minimal value of $D(v)$ may be 0 when
$v$ does not lies in the largest cluster of the network}. $D(v)$ is
1 when the induced subgraph of $V-\{v\}$ is a connected component.
$D(v)=N-1$ when $v$ is the central vertex of a \emph{star-like
network} (that is the graph with $N-1$ vertex of degree 1 being
connected to a central vertex of degree $N-1$). If the degree of
$v$, $d(v)$, is given, a tighter upper bound of $D(v)$ can be given
by:
\begin{equation}N-\frac{N}{d(v)}=N(1-\frac{1}{d(v)})
\end{equation} The upper bound is reachable when
$v$ connects to $d(v)$ clusters that have the same size and
disconnect to each other.

\begin{figure}
\centering
\includegraphics[scale=1]{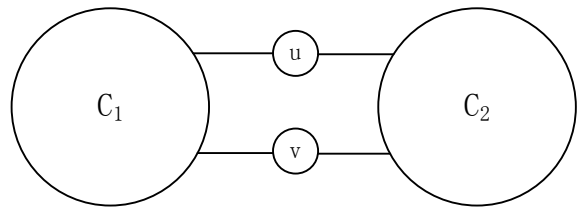}
 \caption{Sensitivity of damage to vertex removal. $C_1$ and $C_2$ are two connected components, each of them has $\frac{N-2}{2}$ vertex.}
\label{fig:syn_damage} 
\end{figure}

Damage of vertices is sensitive to vertex deletion operation. For
example, damages of $u, v$ in the hypothetical graph shown in
Figure~\ref{fig:syn_damage} are both 0. However, once one of $u, v$
is removed from the graph, the damage of the other vertex will
drastically increase approximately to $N/2$. Hence, it is reasonable
to assume that adversaries tend to recalculate damage to maximize
destructive effect.

Let's have a closer look at vertex of different damage values. In a
connected graph, a vertex has a damage value larger than 1 if and
only if it is a \emph{cut} of the graph, which is a vertex whose
removal will increase the number of connected components. The
vulnerability of a network to vertex removal can be attributed to
the existence of these cuts. Vertex of damage 1 are those vertex $v$
such that the induced subgraph of $V-\{v\}$ is connected. The
removal of anyone of these vertex will only isolate itself from the
largest component. Among one-damage vertex, some of them are vertex
of degree 1. Others in one-damage vertex may have diverse degrees.
Since removing an one-damage vertex causes no extra damage to the
network, these vertex all also called \emph{vertex of trivial
damage} in following texts.

Concepts of damage is closely related to \emph{connectivity} of
graphs. Given a connected graph $G$, its vertex-connectivity,
denoted $\kappa_v(G)$ is the minimal number of vertices whose
removal will disconnect $G$ or reduce it to a 1-vertex graph. A
graph $G$ is $k-connected$ if it is connected and the
$\kappa_v(G)\geq k$. Thus, any graph with $\kappa_v(G)\geq 2$ will
not contain vertex with damage larger than 1. In other words, we can
only find vertex of damage larger than 1 from $1-connected$
networks.

\subsection{Damage in synthetic networks}
In this subsection, we will investigate vertex damage in typical
synthetic networks. One is BA networks, accounting for a typical
class of networks with scale-free degree distribution. The second is
ER networks, a typical network model producing exponential degree
distribution. Finally, trees and tree-like networks will also be
investigated.

The connectivity of BA networks depends on the connectivity of it's
seed network and the number of vertex ($m\in N$) that at each step a
newly added vertex connect to. Note that when $m=2$, each step of BA
network generation is a \emph{path addition} to the existing
network, i.e., adding a path with internal vertex disjoint with the
existing vertex to the network. Due to \emph{Whitney Synthesis
Theorem}~\cite{whitney}, the resulting network obtained by
repeatedly performing path addition on a $2-connected$ network is
still $2-connected$. Furthermore, a network is still $2-connected$
if it is obtained by adding edges to a $2-connected$ network. Hence,
for BA models, \emph{if the seed network is $2-connected$ and $m\geq
2$, the resulting network is $2-connected$}. The direct consequence
of this fact is that the BA networks generated as above will not
contain vertex of non-trivial damage value.

An ER random network is generated on $N$ vertices by linking each
pair of vertex with identical probability $0<p<1$. The ER networks
generated under parameters $N$ and $p$ are usually denoted by
$\mathcal{G}(N,p)$. Random graph theory has shown us that \emph{for
$k>0$, almost every graph is k-connected}~\cite{graph}. A graph
property $P$ is said to hold for 'almost every graph' if the
probability that a random graph $G\in \mathcal{G}(N,p)$ has property
$P$ has the limit 1 as $n\rightarrow\infty$. Thus, it's hard to find
vertices with non-trivial damage from a large ER random graph.
However, when $p$ is small and the network is not very large, the
probability that the network is not $k-connected$ is significant.

Let $u,v$ be any two vertices in a graph in $\mathcal{G}(N,p)$, then
other $N-2$ vertex can be partitioned into $\lfloor (N-2)/k \rfloor$
$k$-subsets (a subset of $k$ vertex), with perhaps a fewer vertices
left over. Let $W$ be a $k$-subset. Then, the probability that every
vertex of $W$ is adjacent to $u$ and $v$ (i.e., $W$ is
\emph{fully-adjacent} to $\{u,v\}$) is $p^{2k}$. Hence, the
probability that none of these $k$-subsets is fully-adjacent to
$\{u,v\}$ is
\begin{equation}q=(1-p^{2k})^{\lfloor(\frac{N-2}{k})\rfloor}\label{equ:pro}\end{equation}
Note that, $1-q$ essentially is the probability that for any pair of
vertex $u,v$, there exist $k$ internally-disjoint paths of length 2
to connect $u,v$. A graph $G$ is $k$-connected if and only if for
each pair $u,v$ of vertex there exist at least $k$ internally
disjoint $u-v$ paths (a path with $u,v$ as ends) in
$G$~\cite{graph}. Hence, $1-q$ is a low bound of the probability
that a graph is $k-connected$.

\begin{figure}
\centering {
\includegraphics[scale=0.35]{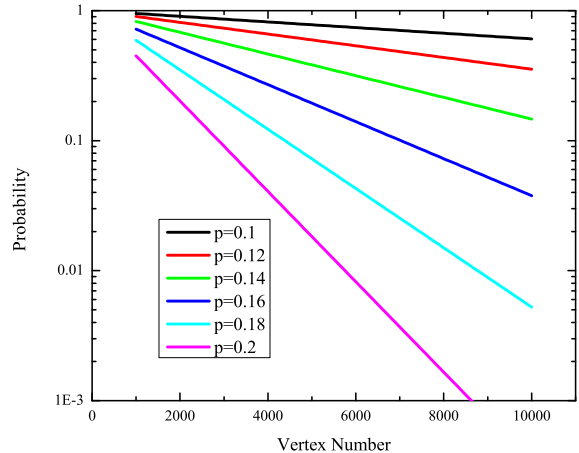}}
 \caption{(Color online) Simulation of probability $q$ in Equation~\ref{equ:pro} with $k=2$.
 $q$ is a function of network size and connectivity probability $p$. In this simulation, we vary the network size from 1000 to
10000 with the increment of 100 to generate 91 samples.}
\label{fig:ER_simu} 
\end{figure}

Thus, $q$ is an upper bound of the probability that a graph is not
$k$-connected. If $k=2$,
$q=(1-p^{4})^{\lfloor(\frac{N-2}{2})\rfloor}$. The simulation of $q$
with $k=2$ as the function of network size and $p$ is shown in
Figure~\ref{fig:ER_simu}. The simulation results show that in ER
network with small size and small $p$, it's still quite possible to
find vertex with nontrivial damage values. For example, when
$N=3000$ and $p=0.18$, $q$ is $20.7\%$, which is significant and can
not be ignored.

\begin{figure}
\centering \subfigure[A path structure]{\label{fig:tree_exa:a}
\includegraphics[scale=0.8]{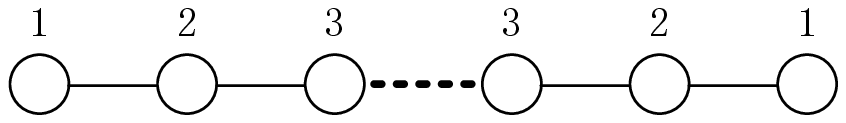}}
\subfigure[A star structure]{\label{fig:tree_exa:b}
\includegraphics[scale=0.4]{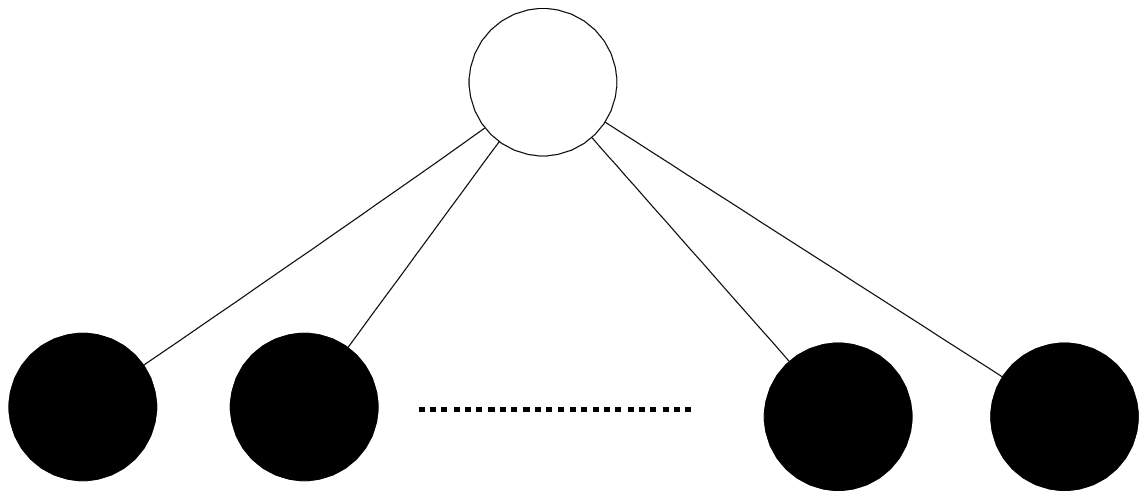}}
 \caption{Two tree structures.}
\label{fig:tree_exa} 
\end{figure}

As connected acyclic graphs, trees are $1-connected$ and vertex with
degrees larger than 1 have nontrivial damage values. The damage
distribution of a tree is determined by it's structure. One extreme
case of tree is path (as shown in Figure~\ref{fig:tree_exa:a}),
where only two ends have degree 1 and all other vertex have degree
2. A path structure has a quite heterogenous damage distribution,
where for each damage value in $[1,\lceil N/2 \rceil]$ there exists
two vertices (when $N$ is odd, only one vertex has damage $\lceil
N/2 \rceil$). The tree that has the most homogenous damage
distribution is a star, as shown in Figure~\ref{fig:tree_exa:b},
where the vertex of largest degree has a extremely large damage
value, that is $N-1$, and all other vertex have damage 1.

\begin{figure}
\centering {
\includegraphics[scale=0.3]{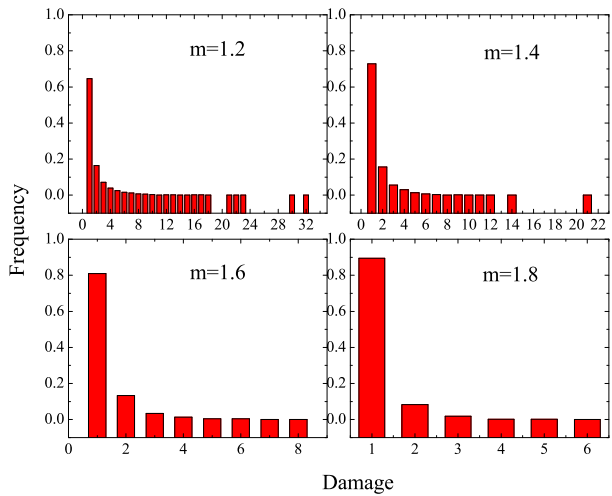}}
 \caption{(Color online) Damage distributions of tree-like networks. The networks are generated by BA network model
 with $m$ set as $1.2,1.4,1.6,1.8$.}
\label{fig:tree_damage} 
\end{figure}

In general, cycles may exist in real-world networks. Hence, it's
hard to find real-world networks taking exactly the form of tree.
However, the structure of a real-world network can be considered as
the union of one of its spanning tree and corresponding additional
edges. In this sense, many networks can be considered as
\emph{tree-like structure}. In general, if the structure of a
network is closer to tree, it's of higher probability to find more
vertex with high damage value. To show this, we give the damage
distribution of four synthetic BA networks with $m$ varying from 1.0
to 1.8 in Figure~\ref{fig:tree_damage}. Note that in the generic BA
network model, $m$ is an integer. To produce treelike BA networks,
we adjust the model to handle cases where $m$ is a decimal in
following ways: each time when a new vertex arrives, we link it to
$\lfloor m \rfloor$ existing vertex with probability $q=m-\lfloor m
\rfloor$ and link it to $\lfloor m \rfloor+1$ existing vertex with
probability $1-q$. When $m=1$, the network is a BA scale-free tree.
When $m$ becomes larger, the network structure is farther away from
the tree structure. It is obvious from Figure~\ref{fig:tree_damage}
that when the network structure is close to tree, long tail can be
observed, indicating that many vertex of large damage exist in the
network.

\subsection{Damage of real networks}
In following texts, we will perform empirical analysis on two real
networks. One is airline transportation network of United States of
year 1997 (denoted by USAir) with vertex representing the airports
of United States and edges representing the airlines. There are
overall 332 airports and 2126 airlines. The other one is
protein-protein interaction network of yeast (denoted by Yeast) with
vertex representing proteins of yeast and edges representing the
interaction of proteins. Yeast has 2361 proteins and 7182
interactions. The behaviors of these two networks under attack are
critical for the functionalities of these networks. The two network
data and their detailed descriptions are now available at
http://vlado.fmf.uni-lj.si/pub/networks/data/.

As shown in Table~\ref{tab:exp_usair}, the airport with the largest
degree in USAir is \emph{Chicago O'hare Intl}, which has 131
airlines connecting to other airports covering $39.5\%$ airports of
US. However, its damage value is only 3, which means that
\emph{Chicago O'hare Intl} is essential for only three airports to
connect to other airports. The airport with the largest damage in
USAir is \emph{Anchorage Intl}, whose damage is 27 implying that 26
airports rely on \emph{Anchorage Intl} as the transferring airport
and \emph{Anchorage Intl} is their unique choice to connect to other
airports. In other words, if \emph{Anchorage Intl} malfunctions,
these 26 airports will be isolated from the outside world if only
air transportation is considered. However, it is surprising to find
that \emph{Anchorage Intl}'s degree is 29 and corresponding degree
rank is only 41. Above analysis clearly shows that the essentiality
of an airport can be characterized by its damage instead of degree.

The top ten airports with the largest degree are shown in
Table~\ref{tab:usair_degree}. Comparing it to
Table~\ref{tab:exp_usair}, we find that only five airports
simultaneously occur in the two ranking lists. It is interesting to
find that some airports that are highly connected, such as
\emph{Chicago O'hare Intl}, \emph{The William B Hartsfield Atlan},
have relatively small damage values. In contrast, some airports with
large damages, such as \emph{Anchorage Intl, Bethel, Honolulu Intl,
Guam Intll, Phoenix Sky Harbor Intl}, have quite small degrees. For
example, \emph{Guam Intll} only connects to 4 airports but has
damage 5, which implies that 4 airports completely rely on
\emph{Guam Intll} to connect to the outside world. Consequently,
despite of its small degree, \emph{Guam Intll} becomes the local
center of air transportation.

Among all the airports, \emph{Dallas/Fort Worth Intl} not only has a
large degree but also has a significant damage. Hence,
\emph{Dallas/Fort Worth Intl} is not only a popular transferring
airports, but also a local center responsible for the connection of
its local airports to the outside world. Above facts together
inspire us that damage has its own right in characterizing the
importance of a vertex from the perspective to maintain the
connections of a vertex's neighborhood to the outside world of the
network.

\begin{table}[t]
\centering
 \caption{\label{tab:exp_usair} Top 10 airports with the largest degrees.}
\begin{tabular}{c c c c}
\hline \hline
Rank & Degree & Damage & Vertex Name\\
\hline

1&  139&    3&  Chicago O'hare Intl \\
2&  118&    15& Dallas/Fort Worth Intl \\
3&  101&    3&  The William B Hartsfield Atlan\\
4&  94& 7& Pittsburgh Intll \\
5&  94& 6& Lambert-St Louis Intl \\
6& 87& 3& Charlotte/Douglas Intl \\
7&  85& 2& Stapleton Intl \\
8&  78& 4& Minneapolis-St Paul Intl/Wold- \\
9&  70& 1& Detroit Metropolitan Wayne Cou \\
10& 68& 7&  San Francisco Intl\\ \hline \hline
\end{tabular}
\end{table}

\begin{table}[t]
\centering
 \caption{\label{tab:usair_degree} Top 10 airports with the largest damages.}
\begin{tabular}{c c c c}
\hline \hline
Rank & Degree & Damage & Vertex Name\\
\hline

 1&  29& 27& Anchorage Intl \\
 2&  118&    15& Dallas/Fort Worth Intl \\
 3&  14& 12& Bethel \\
 4&  94& 7&  Pittsburgh Intll \\
 5&  68& 7&  San Francisco Intl \\
 6&  24& 7&  Honolulu Intl \\
 7&  94& 6&  Lambert-St Louis Intl \\
 8&  4&  5&  Guam Intll \\
 9&  78& 4&  Minneapolis-St Paul Intl/Wold- \\
 10& 60& 4&  Phoenix Sky Harbor Intl\\ \hline \hline
\end{tabular}
\end{table}

\subsection{Correlation between damage and degree}
Degree is the one of the most widely used measures to characterize
the importance of vertex. Whether degree and damage has certain
correlation, for example high degree leading to high damage, is an
intriguing problem. To address issue, we will first study the
correlation between damage and degree for the two real networks
analyzed in previous sections. Then, we propose a network generation
model to produce networks with arbitrary degree-damage correlation.

\begin{figure}
\centering \subfigure[PPI of yeast]{\label{fig:dist:a}
\includegraphics[scale=0.3]{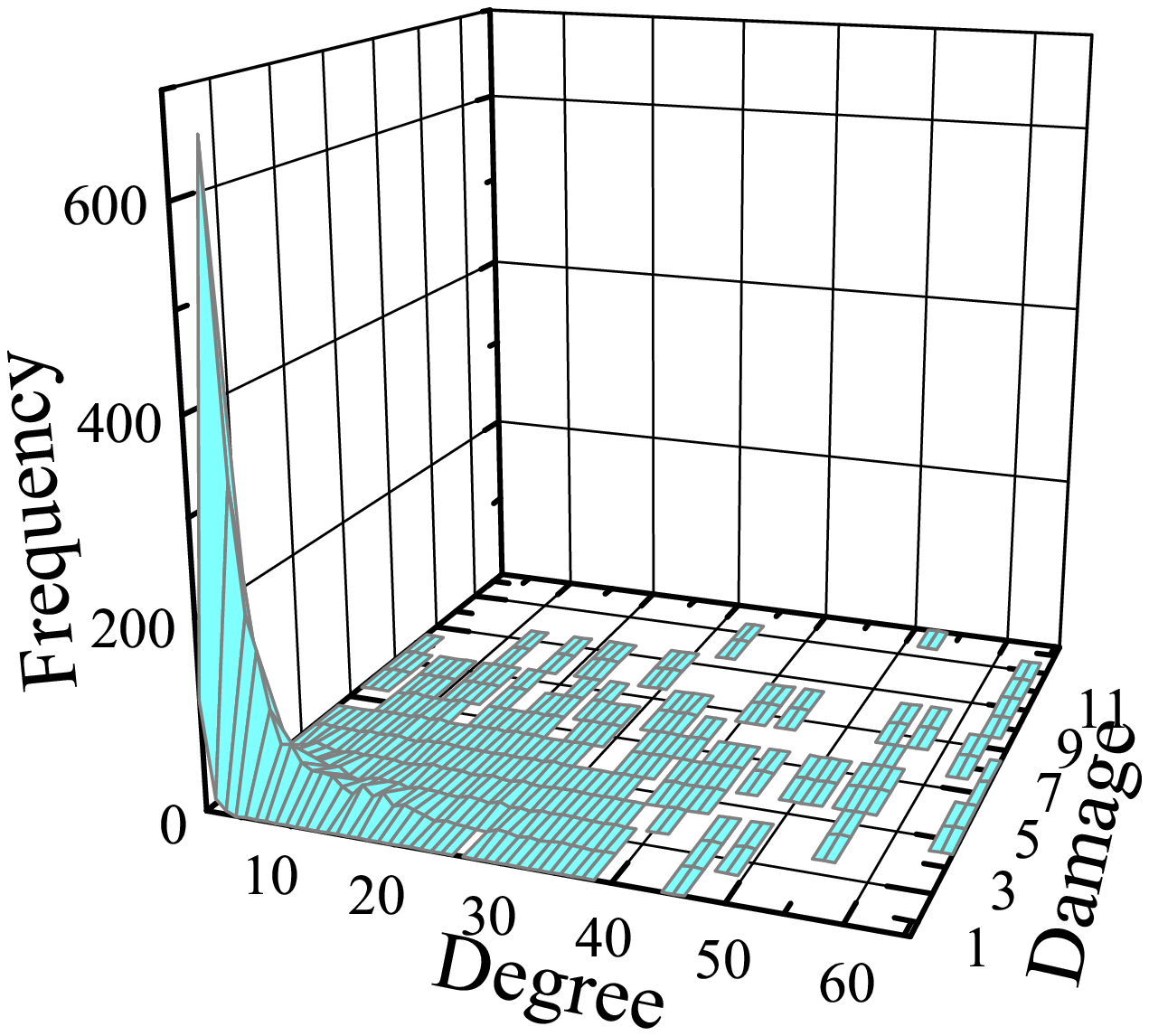}}
\subfigure[Air transportation network in US]{\label{fig:dist:b}
\includegraphics[scale=0.3]{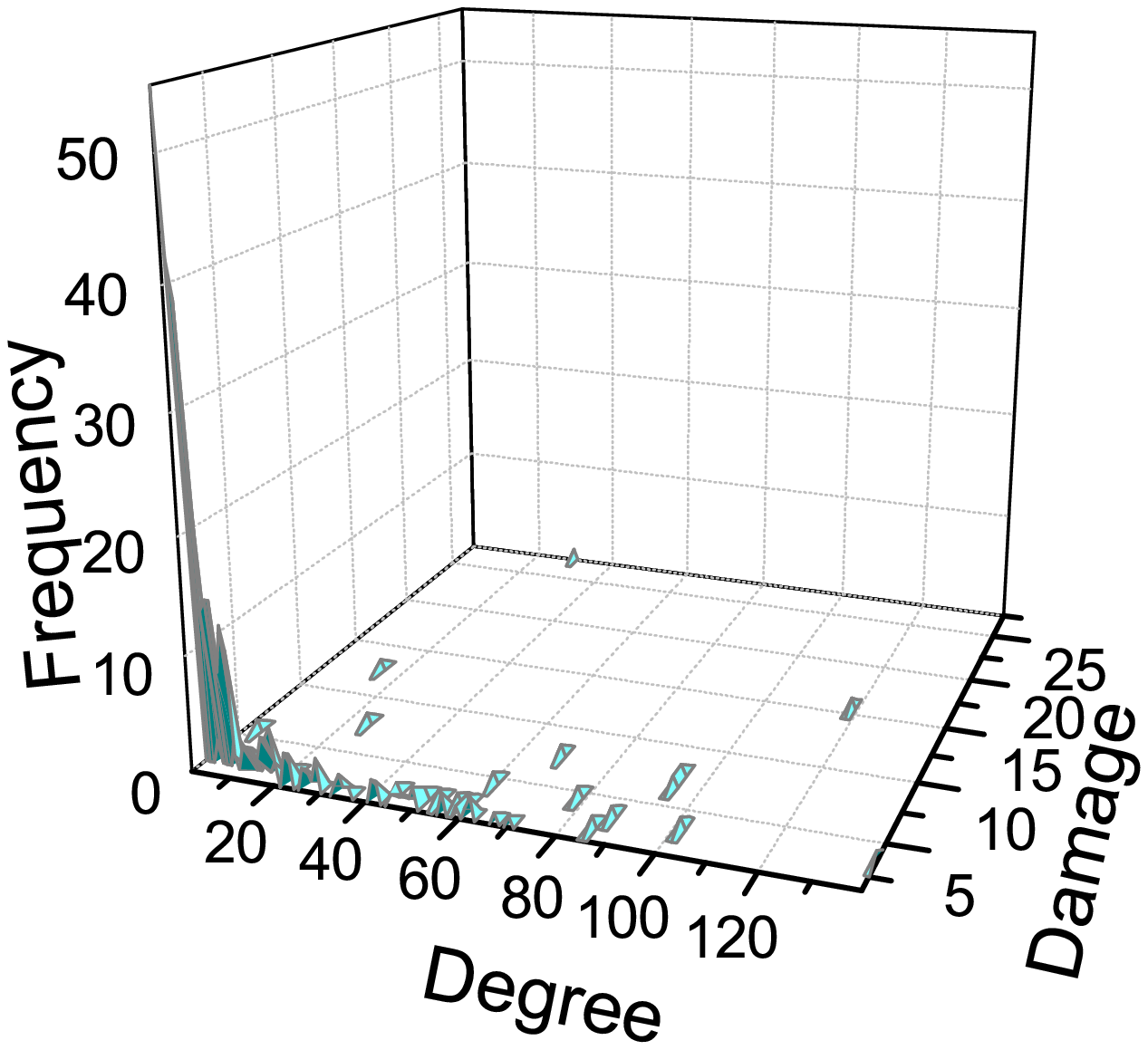}}
 \caption{(Color online) Correlation between degree and damage for Yeast and USAir.}
\label{fig:dist} 
\end{figure}

The distributions of damage and degree for Yeast and USAir are shown
in Figure~\ref{fig:dist}. It is visually apparent that the majority
of vertex tend to have small degrees and small damages. However, in
general, vertex of small degrees do not necessarily have small
damages. Many vertex of small degrees but having larger damage exist
in Yeast, implying that the removal of vertex with small degree may
also have non-ignored influences on the integrity of Yeast.

More generally, there exists no essential correlations between
damage and degree for an arbitrary network, which is verified by a
network generation model that can be tuned to produce networks with
desired correlations between degree and damage. The independence of
damage on degree strongly suggest that damage is a novel perspective
that can not replaced by degree to characterize the property of
vertex.

The network generation model is as follows: Suppose that we need to
generate a network with $N$ vertex. We first partition vertex set
into two subsets $V_1$, $V_2$ with $N_1$ and $N_2$ vertices,
respectively, such that $N=N_1+N_2$. Then, we generate a graph
(denoted by $G[V_1]$) with vertex set $V_1$ such that any vertex in
$G[V_1]$ has damage 1. This objective can be achieved if $G[V_1]$ is
$k-connected$ and $k\geq 2$. BA network generator that uses a
complete subgraph as the seed network and at each step a newly added
vertex is attached to $m\geq 2$ existing vertices can help us
generate a $2-connected$ $G[V_1]$.

\begin{figure}
\centering {
\includegraphics[scale=0.32]{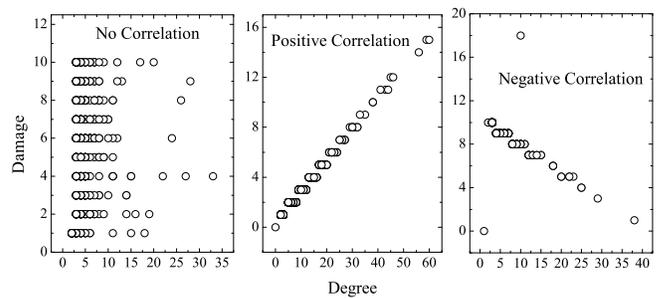}}
 \caption{Synthetic networks with desired correlation between damage and degree. Each generated networks consists of 2000 vertex and 3500 edges with $N_1=1500$ and $N_2=500$.
 $G[V_1]$ is generated by BA model with $m=2$.}
\label{fig:corr} 
\end{figure}

In the second step, for each vertex $u$ in $V_2$, we will attach it to a vertex $v_i\in V_1$ of degree $k_i$ by probability
\begin{equation}
\Pi(v_i)\sim \frac{f(k_i)}{\sum_{v_i\in V_1}f(k_i)}
\end{equation}, where $k_i$ is degree of $v_i$ in $G[V_1]$ and $f(k_i)$ is a function of $k_i$.
In the resulting graph, any vertex $v_i$ in $V_1$ has degree
$k_i+\Pi(v_i)N_2$ and damage $\Pi(v_i)N_2+1$, while each vertex of
$V_2$ has degree 1 and damage 1. Consequently, we can generate a
network stratifying desired correlation between damage and degree by
tuning $f(k_i)$. As an example, we can generate graphs such that its
correlation between degree and damage is (1)Positive, (2)Negative,
(3)Independent by specifying $f(k_i)$ as:
\begin{enumerate}
\item $f(k_i)\sim k_i$;
\item $f(k_i)\sim k_i^{-1}$;
\item $f(k_i)\sim k_i \mod \mu$, where $\mu$ is a large prime
\end{enumerate}, respectively.
The correlation plots of networks generated by above parameters are
shown in FIG~\ref{fig:corr}. The figure shows that the network model
can generate networks with different correlations between degree and
damage.

\section{Damage attack}
Now, we are ready to give the detailed procedure of damage attack:
\emph{An informed agent always attempts to deliberately attack a
vertex with the maximal damage value in the current network. After
the malfunctioning of the target vertex, simulated by the removal of
the vertex, the agent recalculates the damage value of each vertex
in the network. The attack continues until the destructive objective
is achieved.} To comprehensively understand the vulnerability of
network under intentional attacks, it is reasonable to assume that
the attack will stop until the network completely falls apart, i.e.,
the largest connected component of the network contains only one
isolated vertex. Compared to degree attack, there are two
distinctive features in damage attack. First, vertex importance is
measured by damage rather than degree. Second, recalculation is
indispensable due to the sensitivity of damage value to vertex
removal.

\subsection{Damage attack on real networks}
We first show the destructive effect of damage attack on a small
hypothetical network with the comparison to degree attack. For the
graph shown in Figure~\ref{fig:example_net}, when the vertex with
the largest damage, i.e., $u_1$, is removed, $S_{max}$ will fast
decrease from 18 to 13. Compared to damage attack, if the vertex
with the largest degree, i.e., $v_1$, is removed, other vertex
remain connected. When the damage attack continues, $u_2$ will be
removed, which will produce a network with only 9 vertex (half of
the original size). However, under degree attack, the target after
the removal of $v_1$ is $v_2$, whose removal has no effect on the
connectedness of the remaining 16 vertex.

\begin{figure}
\centering \subfigure[Yeast]{\label{fig:real:a}
\includegraphics[scale=0.3]{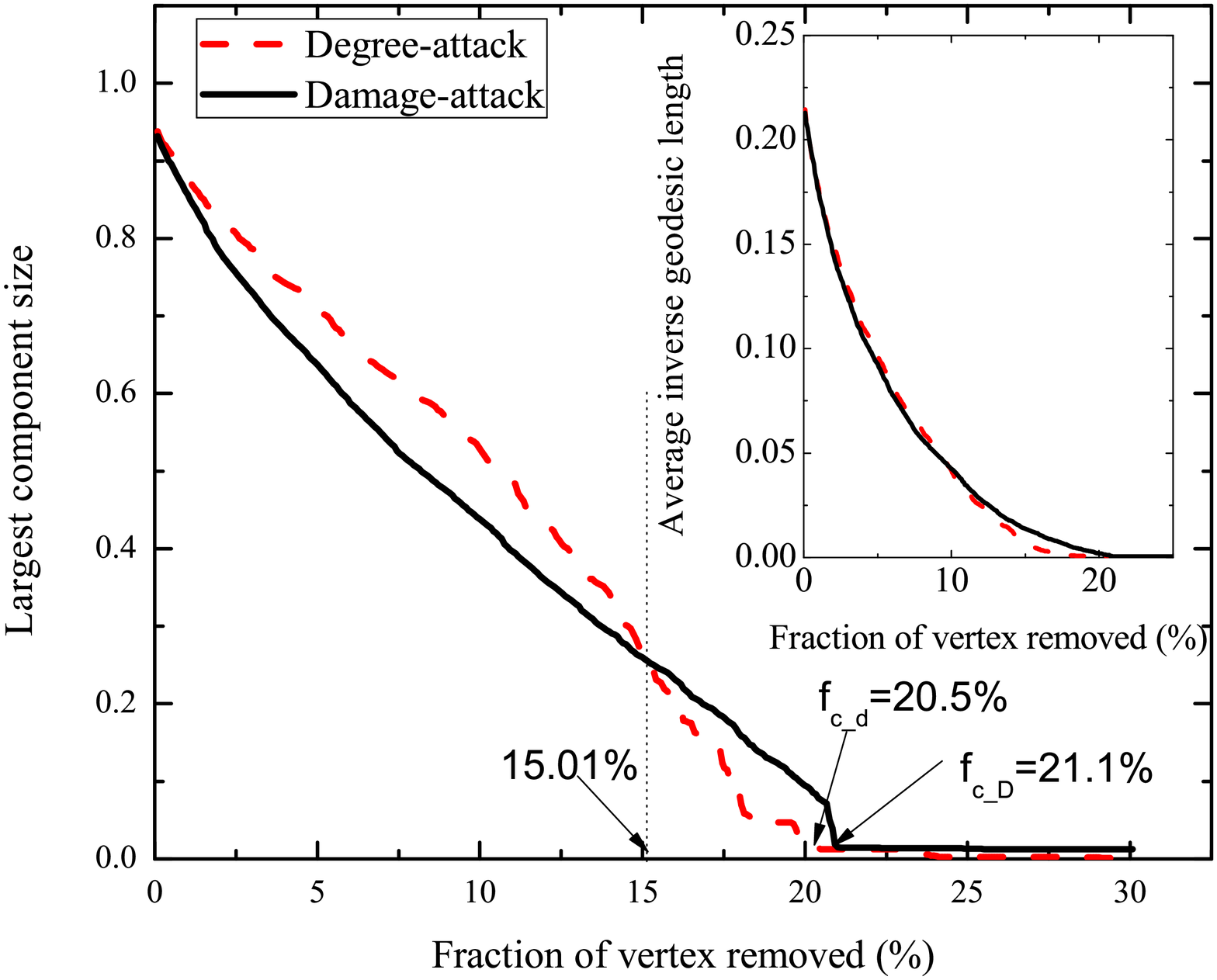}}
\subfigure[USAir]{\label{fig:real:b}
\includegraphics[scale=0.3]{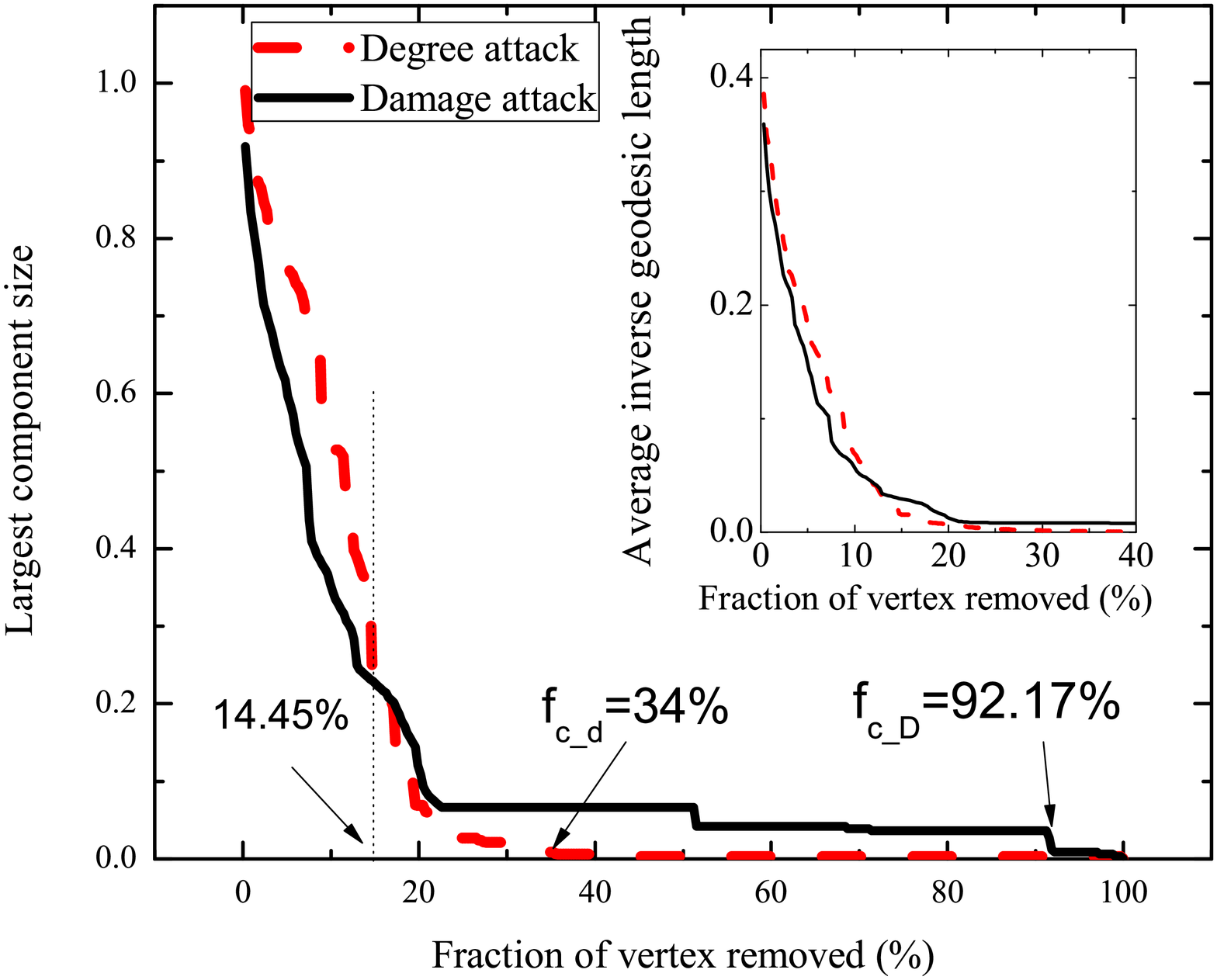}}
 \caption{(Color online) Intentional attacks guided by damage and degree on real networks.
The insets are the results of degree attack and damage attack when
network performance is measured by average inverse geodesic length
$l^{-1}$ as defined in Eq.(\ref{equ:spi}).}
\label{fig:real} 
\end{figure}

Above example shows that \emph{damage attack is more destructive
than degree attack, which is valid at least at the early stage of
the attack}. Then, we may wonder \emph{whether the superiority of
damage attack is consistent across the whole attacking process until
the network completely falls apart}. To answer this question, we
simulate the intentional attack guided by damage and degree,
respectively, on two real networks used in previous sections. The
result is shown in Figure~\ref{fig:real}. It is visually apparent
from the plots that for both the two real networks the destruction
(measured by the decrease of $S_{max}$ and average inverse geodesic
length $l^{-1}$) caused by damage attack is more significant than
that caused by degree attack until a cross-point is reached. Such
facts sufficiently show that \emph{damage attack is more destructive
than degree attack before the cross-point}.

Specifically, for Yeast, before the fraction of vertex removed
reaches to the cross-point 15.01\%, $S_{max}$ under damage attack is
always smaller than that under degree attack. However, after 15.01\%
degree attack shows more efficiency than damage attack until the
network completely collapses. It is approximately at
$f_{c_d}=20.5\%$ (under degree attack ) and $f_{c_D}=21.1\%$ (under
damage attack) that the network completely falls apart. If complete
destruction is the attacking objective, degree attack shows minor
superiority to damage attack, since $f_{c_D}$ is a little larger
than $f_{c_d}$.

Note that at the cross-point $15.01\%$, $S_{max}$ is less than
$25\%$ of the original size, which implies that at the cross-point
Yeast generally has already lost most of its functionalities. Before
the cross-point, when fraction of removed vertex reaches to
approximately $8.5\%$, maximal extra destruction caused by damage
attack compared to degree attack is reached, which is about $9.7\%$
of original network size. The extra damage is significant enough,
since in many real networks, the network almost lost its function
when removing $9.7\%$ vertex. In Yeast, superiority of damage attack
dominates the whole attacking process since at the cross-point, 3/4
attacking process has finished. All these facts together suggest
that destruction caused by damage attack is more significant and can
not be ignored in real applications. The fragility of Yeast under
damage attack also provides additional evidence for the notion that
damage characterizes the essentiality of proteins in PPI
networks~\cite{dam-ppi}.

Similar phenomenon can be observed from USAir network. The results
are shown in Figure~\ref{fig:real:b}. Before the cross-point
$14.45\%$, damage attack is more destructive than degree attack.
This result is consistent when the network performance is measured
by average geodesic length, as shown in the inset of
Fig.~\ref{fig:real:b}. When cross-point is reached, $S_{max}$ is
only about $20\%$ of the original size, suggesting that network has
been fragmented into pieces. It is surprising to find that the
maximal extra destruction of damage attack is $27.96\%$, which is
obtained when fraction of removed vertex reaches to $8.7\%$. These
facts strongly suggest that USAir network exhibits more fragility
under damage attack than under degree attack. Note that in USAir
$f_{c_D}$, approximately to be $92.17\%$, is quite large compared to
$f_{c_d}=34\%$. However, these indicators are meaningless since
after the cross-point the attack is performed on a network with
disconnected pieces (note that $S_{max}$ after cross-point is less
than $20\%$ of the original network size).

\subsection{Damage attack on synthetic networks}
Next, we will show that previous observations on real networks are
consistent on synthetic networks including ER networks and BA
scale-free networks. Most of real networks can be reproduced by
these two synthetic network models under suitable parameters. In
this subsection, we generated an ensemble of BA and ER networks (10
realizations, respectively) with the same parameters, so that each
performance quantity can be evaluated as the average of 10
realizations.

\begin{figure}
\centering \subfigure[BA Network]{\label{fig:syn:a}
\includegraphics[scale=0.3]{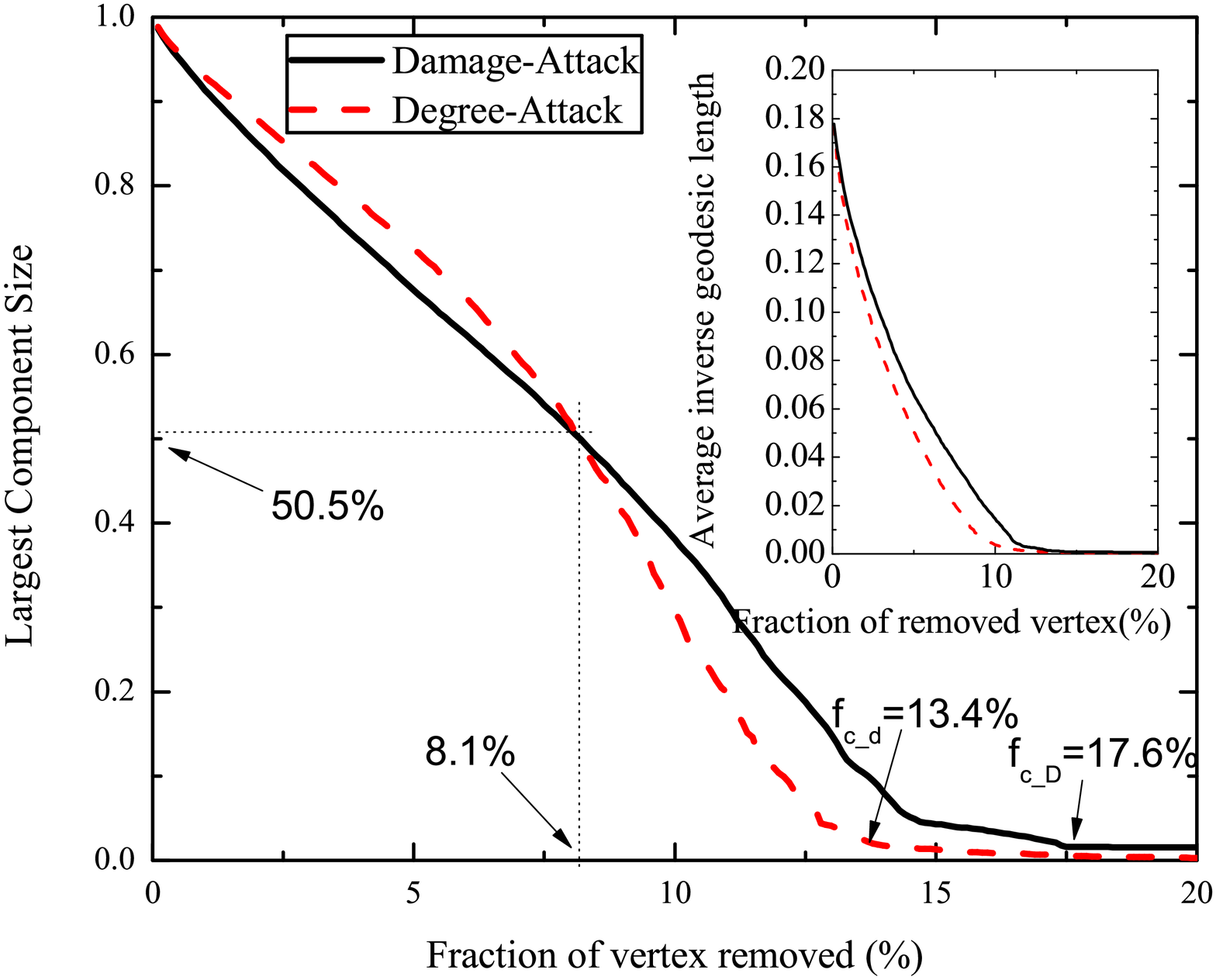}}
\subfigure[ER network]{
\includegraphics[scale=0.3]{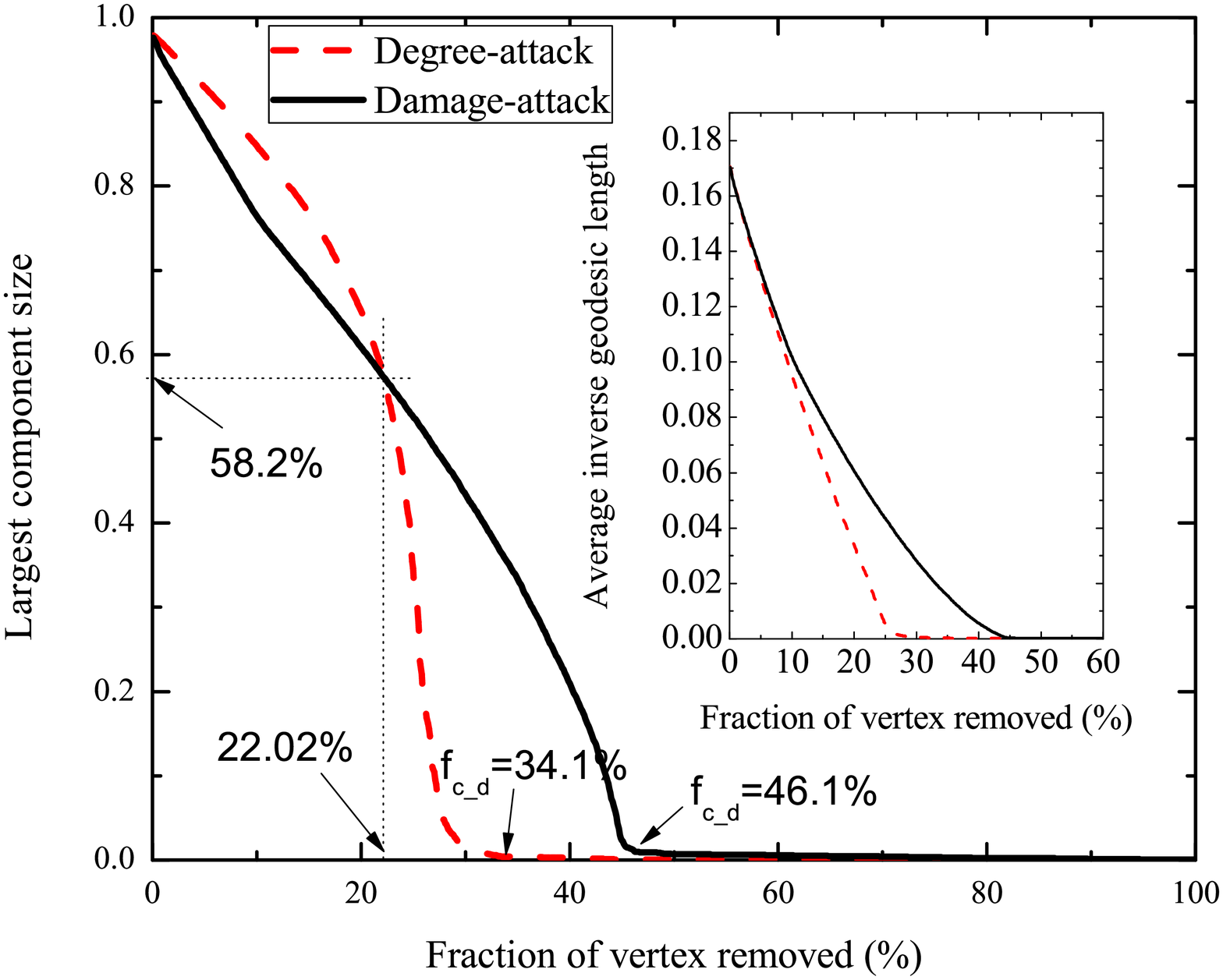}}
 \caption{(Color online) Intentional attacks guided by damage and degree on synthetic networks.
 Each value in the figures is evaluated by the
 average on 10 realizations of BA (ER) networks with the same
 parameters. The variance of the 10 samples is close to 0, thus being omitted in the figure.
 Figure \ref{fig:syn:a} shows the
 result on BA scale free networks with $N=3000$ and average degree $<k>=3.1$.
  Figure \ref{fig:syn:a} shows the
 result on ER networks with $N=3000$ nodes and $<k>=4$.}
\label{fig:syn} 
\end{figure}

The result on BA network is shown in Fig.~\ref{fig:syn:a}. Existence
of a cross-point (8.1\%) of damage-attack curve and degree-attack
curve is obvious from the figure. Damage attack still exhibits more
destructive effect on BA networks than degree attack before the
cross-point. It is about at $5\%$ that the maximal extra destruction
caused by damage ($\approx 5\%$) can be reached before the
cross-point. After the cross-point, $S_{max}$ under degree attack
drops faster than that under damage attack, yielding a small
$f_{c\_d}$ ($\approx 14.3\%$) and relatively large $f_{c\_D}$
($\approx 17.6\%$). However, when using $l^{-1}$ as the measure of
network performance, it seems that degree attack is more destructive
than degree attack along the entire attacking process (as shown in
the inset of Fig.~\ref{fig:syn:a}).

Similar results can be observed from ER networks, with following
distinctive observations. First, for ER networks, the cross-point
($22.2\%$) and the critical points when the network completely falls
apart ($f_{c_d}\approx 34.1\%$ and $f_{c_D}\approx 46.1\%$) under
two attacks are significantly larger than corresponding counterparts
of BA network. Such facts suggest that ER networks are more robust
against intentional attacks including degree attack and damage
attack. In general, the integrity of ER network is maintained by a
majority of vertex with average degree. In contrast, the integrity
of BA network heavily relies on the minority of vertex with largest
degrees. Hence, when adversaries deliberately attack the vertex with
the largest degree or damage in the network, BA network is more
fragile than ER networks.

 Comparing the results on real networks and
synthetic networks, we find that damage attack seems to be more
effective on real networks than on synthetic networks. At the
cross-point the destructive effects on real networks are more
significant than that on synthetic networks (both ER network and BA
network). For real networks, $S_{max}$ at the cross-point lies in
the range $(20\%, 30\%)$; whereas for synthetic networks, $S_{max}$
varies from $50\%$ to $60\%$. In reality, when only 20\%~30\% vertex
remain in the largest connected component, the function of the whole
system is reasonably to be regarded as collapsed. Hence, damage
attack is really effective to destruct real networks.

\subsection{Summary}
Comparatively empirical studies of damage attacks and degree attacks
on both real networks and synthetic networks strongly suggest that:
\begin{enumerate}
\item Real networks and typical synthetic networks are quite vulnerable to damage attack;
\item The existence of a cross-point is a universe phenomenon shared across a variety of diverse networks, when comparing degree attack and damage attack with network functionality being measured by $S_{max}$;
\item Damage attack is more destructive than degree attack before the cross-point, at which a network generally has already lost most of its functionalities.
\end{enumerate}
All these findings together show that damage attack poses a great
challenge for us to protect complex networks, hence deserving our
research efforts.

\section{Analysis}
Findings in the previous sections are not self-evident. In this
section, we will explore the reasonable explanations for above
results.

Note that, after the cross-point of degree attack and damage attack,
the decrease of $S_{max}$ under degree attack is faster than that
under damage attack. For example, for BA networks, under degree
attack, 50.5\% decrease of $S_{max}$ is caused by the removal of
13.4\%-8.1\%=5.3\% vertex, while under damage attack the same
destruction is caused by the removal of 17.6\%-8.1\%=9.5\% vertex.
Such facts imply that the integrity of the remaining network under
degree attack at the cross-point is maintained by only a minority of
vertex, whose removal will cause fast collapse of the network. Then,
it's reasonable to expect that the network at the cross-point under
degree attack is more sensitive to the removal of some key vertex
than that under damage attack.

\begin{figure}
\centering \subfigure[Damage distributions of
Yeast]{\label{fig:damage_dist:a}
\includegraphics[scale=0.3]{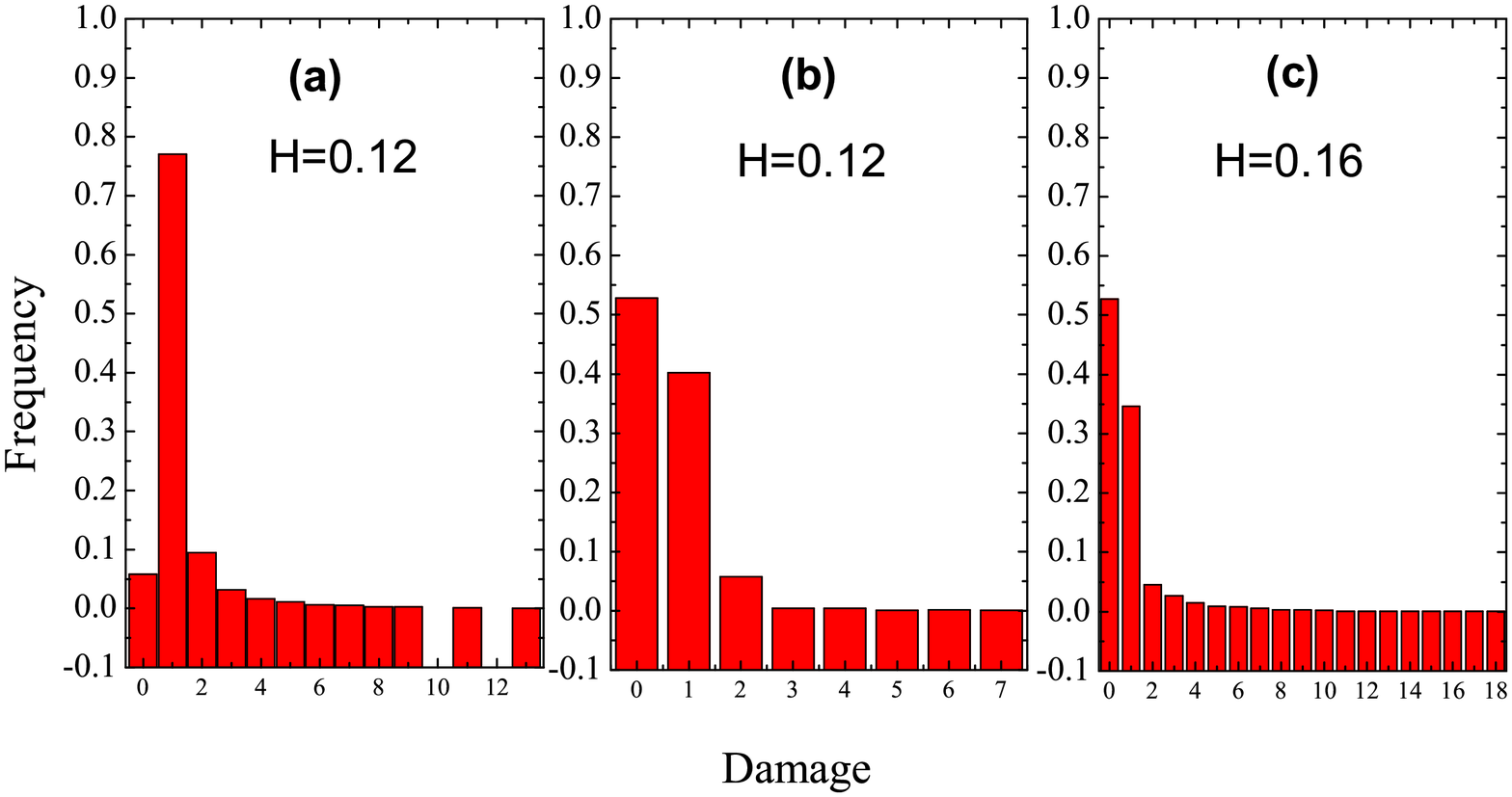}}
\subfigure[Damage distributions of USAir]{\label{fig:damage_dist:b}
\includegraphics[scale=0.3]{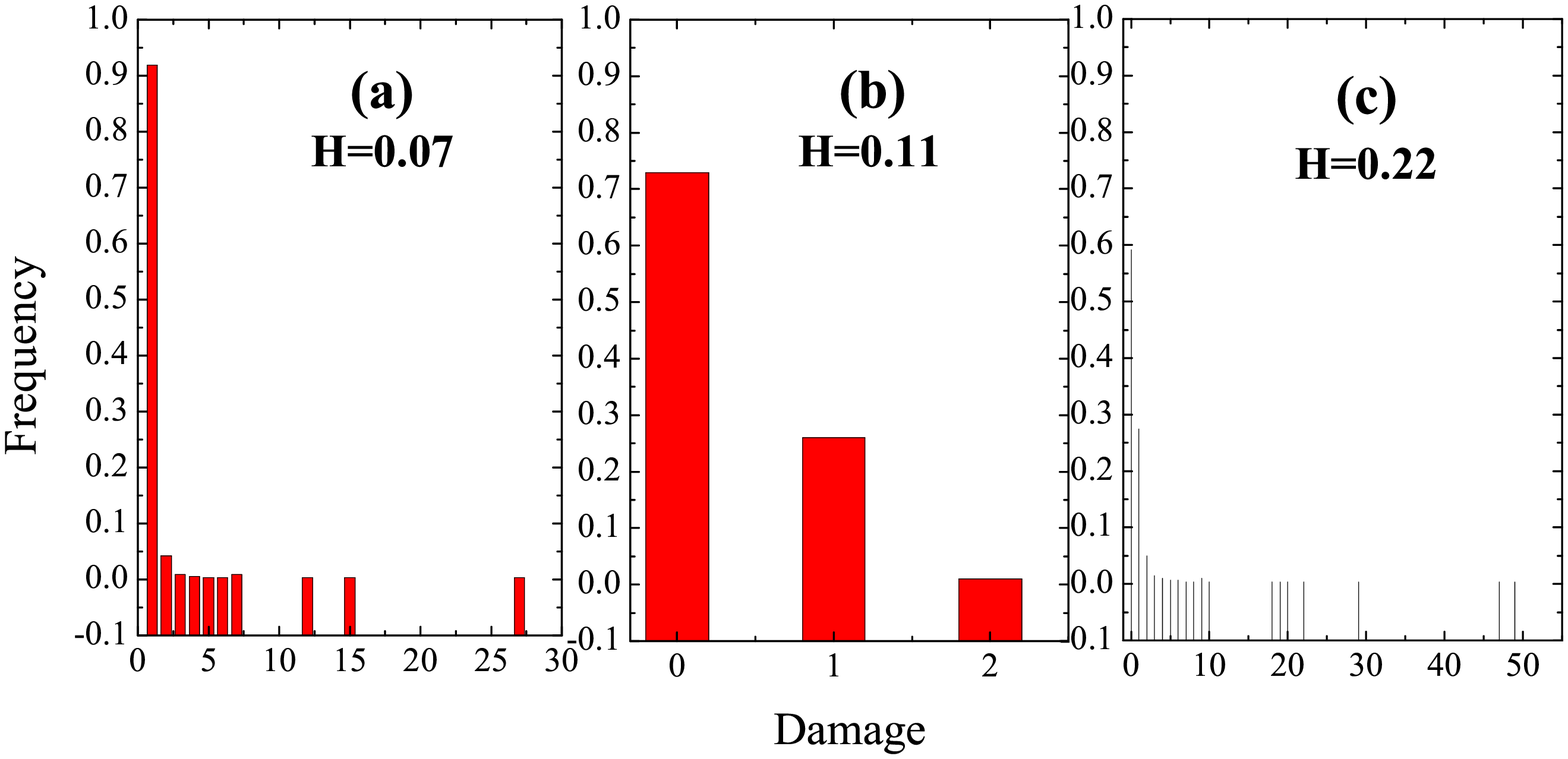}}
 \caption{(Color online) Damage distributions of two real networks. (a) shows the damage
 distributions of the original networks. (b) and (c) show the damage distribution of $G_D$ and $G_d$, respectively. $H$ in each figure
 is the entropy value computed by Eq.~(\ref{equ:entropy}). }
\label{fig:damage_dist} 
\end{figure}

The sensitivity to vertex removals can be observed from the damage
distributions of the remaining networks. Let $G_d, G_D$ be the
networks at the cross point under degree attack and damage attack,
respectively. We summarize the damage distributions of $G_d,G_D$ for
Yeast and USAir in Figure~\ref{fig:damage_dist}. As comparisons, the
original damage distributions are also given. Let $P(D)$, $P_d(D)$
and $P_D(D)$ be the damage distribution of the original network,
$G_d$ and $G_D$, respectively. From Figure~\ref{fig:damage_dist}, we
can see that for both two real networks $P_d(D)$ is more
right-skewed than $P(D)$ while $P_D(D)$ is less right-skewed than
$P(D)$, suggesting that $P_D(D)$ is more homogenous than $P(D)$ and
$P_d(D)$ is more heterogenous than $P(D)$. Consequently, $G_d$ is
more sensitive to degree attack than $G_D$ to damage attack.

To give quantitative support, we use entropy to measure the
heterogeneity of a damage distribution. The entropy is defined as:
\begin{equation}H(G)=-\sum p_i\log{p_i}
\label{equ:entropy}
\end{equation}, where $p_i$ is the probability that a vertex has damage
$i$. The larger $H(G)$ is the more heterogeneous the distribution
is. The entropy values are also given in
Figure~\ref{fig:damage_dist}, which confirm that damage distribution
of $G_d$ is more heterogeneous than $G_D$.

Now let's have a closer look at the damage distributions at the
cross-point. For USAir, two airports \emph{'Fort
Lauderdale/Hollywood Intl'} and \emph{'General Mitchell Intll'} have
damage value as 47, 49, respectively, in $G_d$. Note that the damage
caused by these vertex are significant, since the airports isolated
from the largest cluster of $G_d$ is approximately $15\%$ of all
airports. Hence, the integrity of $G_d$ heavily relies on the
existence of these two airports. However, it is surprising to find
that both of these two airports have only damage as 1 and degree as
34 in the original network. This fact suggests that those vertex of
less importance (whatever quantified by degree or damage) in the
original network may become the most important vertex in the
remaining networks after a number of steps of attacks guided by
degrees. Hence, \emph{one of the most important characteristics of
degree attack is the emergence of potential important vertex.}
However, for damage attack, the case is just the reverse: \emph{as
the attack continues, more vertex tend to have similar damage
values, and consequently damage attack gradually degrades into
random attack}.

\begin{figure}
\centering \subfigure[Damage distributions of two
snapshots.]{\label{fig:damage:a}
\includegraphics[scale=0.3]{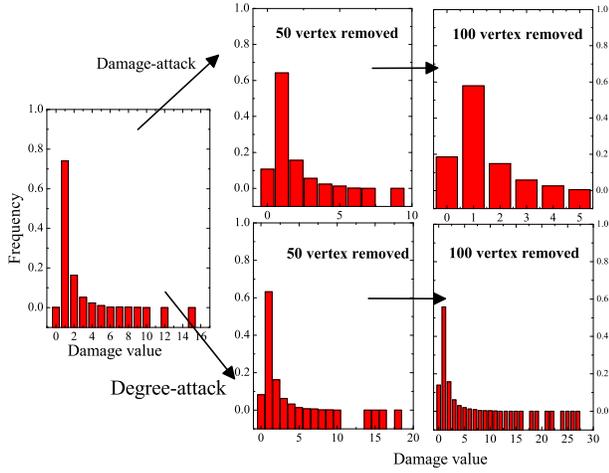}}
\subfigure[Entropies of damage distributions as the function of the
number of vertex removed.]{\label{fig:damage:b}
\includegraphics[scale=0.3]{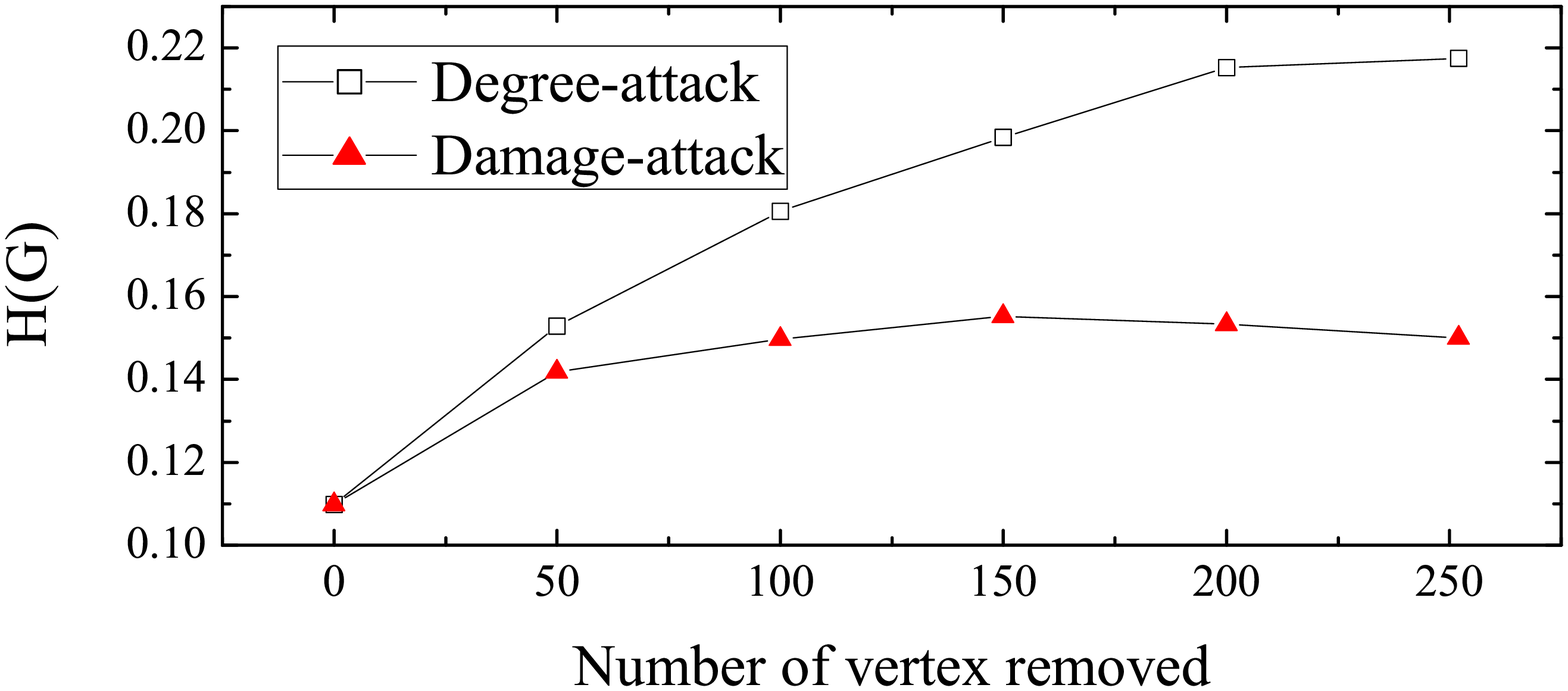}}
 \caption{(Color online) Evolution of damage distributions under degree attack and damage attack.
(a) compares the damage distributions when 50, 100 vertex removed
under degree attack and damage attack. (b) shows the entropy of
damage distribution as the function of number of removed vertex
under two attack strategies. The statistics is summarized from
 a BA network with 3000 vertex and 4369 edges.}
\label{fig:damage} 
\end{figure}

To provide more quantitative evidences, we further investigate the
evolution of damage distribution under degree attack and damage
attack. We capture six snapshots of the intermediate results under
degree attack and damage attack from a BA network (similar results
can be obtained from real networks tested in above sections and ER
networks). The result is shown in Figure~\ref{fig:damage:b}, which
clearly shows that across the whole attacking process the damage
distribution under degree attack is consistently more heterogenous
than that under damage attack. It seems that under damage attack,
damage distribution converges to a constant level. To have a direct
feeling about the heterogeneity, the damage distributions of three
snapshots in Figure~\ref{fig:damage:b} are also given in
Figure~\ref{fig:damage:a}, which are consistent with the above
results.

Now, it's ready to explain why degree attack produces networks with
more heterogenous damage distribution than damage attack. Note that,
under degree attack, more edges will be removed from the network
compared to damage attack when the same number of vertex are
removed. The consequence of this is the sparsity of the resulting
network. Compared to degree attack, damage attack seems to be more
destructive at the early attacking stage since more vertices are
isolated from the largest cluster, however the number of edges
isolated from the largest cluster generally is less than degree
attack. Consequently, when it comes to a critical point where the
accumulative effect of degree attack becomes significant, the
integrity of the network will collapse avalanchely.

\begin{figure}
\centering {
\includegraphics[scale=0.3]{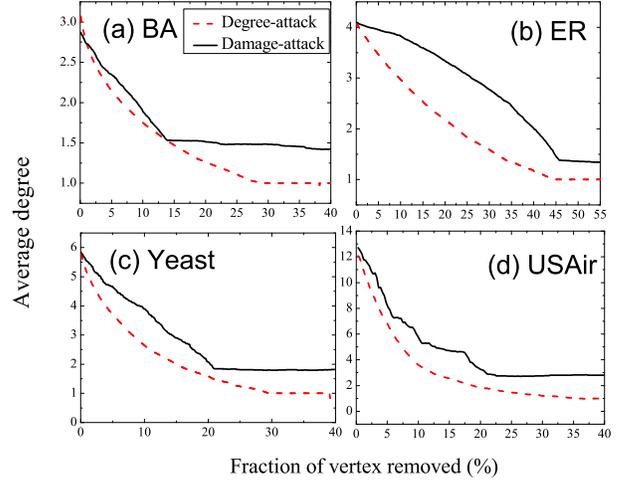}}
 \caption{(Color online) Average degree as the function of fraction of vertex removed. The synthetic networks are the same as that used in Figure~\ref{fig:syn}.
 The result of synthetic networks is summarized as the average of 10 realizations with the same parameters as previous experiments.}
\label{fig:avg_degree} 
\end{figure}

To verify above conjectures, we summarize the average degrees (the
ratio of edge number to vertex number) for intermediate
networks~\footnote{Note that the intermediate network may contain
disconnected components. Hence, the average degree is summarized
from the entire intermediate network instead of its largest
connected component.} under degree attack and damage attack. The
results of two synthetic networks and two real networks are shown in
Figure~\ref{fig:avg_degree}. For all tested networks, under degree
attack, the average degree of an intermediate network is
consistently smaller than that under damage attack, indicating that
degree attack is more destructive in removing edges from a network.

\section{Conclusions}
In this paper, we first review existing attacking models with the
objective to unify existing models. Then, we systematically
investigate damage and its distributions in two typical real
networks (USAir and PPI of Yeast) and typical synthetic networks
including BA networks, ER networks and tree-like networks. We show
that BA network generated from a 2-connected seed network with
$m\geq 2$ is also a 2-connected network. We also show that vertex of
higher damage tend to exist in a tree-like network. Statistics about
damages in two networks show that damage has its own right in
characterizing the importance of a vertex in maintaining the
connection of its neighborhood to the outside world of the network,
which further suggests that as a measure of vertex, damage can not
be trivially replaced by degree or other measures.

We empirically analyze the behaviors of complex networks against
damage attack with the comparisons to degree attack. Our major
finding include: There exists a cross-point of degree attack and
damage attack before which damage attack is more destructive than
degree attack for a variety of diverse networks; Real networks and
typical synthetic networks are quite vulnerable to damage attack
since at the cross-point the network already lost most of its
functionalities. Further investigation shows that degree attack
tends to produce networks with more heterogenous damage distribution
than damage attack, which accounts for the existence of the
cross-point.

All above findings together suggest that damage attack is one of
most potentially destructive attacks, deserving further research
efforts. The vulnerability of real networks and synthetic networks
also poses a great challenge for us to protect these networks.
Results in this paper may shed light on efficient solutions to
protect complex networks against damage attack.

\bigskip
{\bf Acknowledgements} This work was supported by the National
Natural Science Foundation of China under grants No.61003001,
No.60703093; Specialized Research Fund for the Doctoral Program of
Higher Education No.20100071120032; Innovation Program of Shanghai
Municipal Education Commission under Grant No.10YS163; the
Specialized Foundation of Shanghai Education Commission for
Outstanding Young Teachers in University under Grant No.sdl-09015.

\section{\label{sec:level2}References and notes}

\bibliography{robust}

\begin{thebibliography}{26}
\expandafter\ifx\csname natexlab\endcsname\relax\def\natexlab#1{#1}\fi
\expandafter\ifx\csname bibnamefont\endcsname\relax
  \def\bibnamefont#1{#1}\fi
\expandafter\ifx\csname bibfnamefont\endcsname\relax
  \def\bibfnamefont#1{#1}\fi
\expandafter\ifx\csname citenamefont\endcsname\relax
  \def\citenamefont#1{#1}\fi
\expandafter\ifx\csname url\endcsname\relax
  \def\url#1{\texttt{#1}}\fi
\expandafter\ifx\csname urlprefix\endcsname\relax\def\urlprefix{URL }\fi
\providecommand{\bibinfo}[2]{#2}
\providecommand{\eprint}[2][]{\url{#2}}

\bibitem[{\citenamefont{Cohen et~al.}(2000)\citenamefont{Cohen, Erez, ben
  Avraham, and Havlin}}]{cohen00}
\bibinfo{author}{\bibfnamefont{R.}~\bibnamefont{Cohen}},
  \bibinfo{author}{\bibfnamefont{K.}~\bibnamefont{Erez}},
  \bibinfo{author}{\bibfnamefont{D.}~\bibnamefont{ben Avraham}},
  \bibnamefont{and} \bibinfo{author}{\bibfnamefont{S.}~\bibnamefont{Havlin}},
  \bibinfo{journal}{Physical Review Letters} \textbf{\bibinfo{volume}{85}},
  \bibinfo{pages}{4626} (\bibinfo{year}{2000}).

\bibitem[{\citenamefont{Cohen et~al.}(2001)\citenamefont{Cohen, Erez, ben
  Avraham, and Havlin}}]{cohen01}
\bibinfo{author}{\bibfnamefont{R.}~\bibnamefont{Cohen}},
  \bibinfo{author}{\bibfnamefont{K.}~\bibnamefont{Erez}},
  \bibinfo{author}{\bibfnamefont{D.}~\bibnamefont{ben Avraham}},
  \bibnamefont{and} \bibinfo{author}{\bibfnamefont{S.}~\bibnamefont{Havlin}},
  \bibinfo{journal}{Physical Review Letters} \textbf{\bibinfo{volume}{86}},
  \bibinfo{pages}{3682} (\bibinfo{year}{2001}).

\bibitem[{\citenamefont{Paul et~al.}(2004)\citenamefont{Paul, Tanizawa, Havlin,
  and Stanley}}]{paul}
\bibinfo{author}{\bibfnamefont{G.}~\bibnamefont{Paul}},
  \bibinfo{author}{\bibfnamefont{T.}~\bibnamefont{Tanizawa}},
  \bibinfo{author}{\bibfnamefont{S.}~\bibnamefont{Havlin}}, \bibnamefont{and}
  \bibinfo{author}{\bibfnamefont{H.}~\bibnamefont{Stanley}},
  \bibinfo{journal}{The European Physical Journal B}
  \textbf{\bibinfo{volume}{38}}, \bibinfo{pages}{187} (\bibinfo{year}{2004}).

\bibitem[{\citenamefont{Tanizawa et~al.}(2005)\citenamefont{Tanizawa, Paul,
  Cohen, Havlin, and Stanley}}]{tani05}
\bibinfo{author}{\bibfnamefont{T.}~\bibnamefont{Tanizawa}},
  \bibinfo{author}{\bibfnamefont{G.}~\bibnamefont{Paul}},
  \bibinfo{author}{\bibfnamefont{R.}~\bibnamefont{Cohen}},
  \bibinfo{author}{\bibfnamefont{S.}~\bibnamefont{Havlin}}, \bibnamefont{and}
  \bibinfo{author}{\bibfnamefont{H.~E.} \bibnamefont{Stanley}},
  \bibinfo{journal}{Physical Review E} \textbf{\bibinfo{volume}{71}},
  \bibinfo{pages}{047101} (\bibinfo{year}{2005}).

\bibitem[{\citenamefont{Tanizawa et~al.}(2006)\citenamefont{Tanizawa, Paul,
  Havlin, and Stanley}}]{tani06}
\bibinfo{author}{\bibfnamefont{T.}~\bibnamefont{Tanizawa}},
  \bibinfo{author}{\bibfnamefont{G.}~\bibnamefont{Paul}},
  \bibinfo{author}{\bibfnamefont{S.}~\bibnamefont{Havlin}}, \bibnamefont{and}
  \bibinfo{author}{\bibfnamefont{H.~E.} \bibnamefont{Stanley}},
  \bibinfo{journal}{Physical Review E} \textbf{\bibinfo{volume}{74}},
  \bibinfo{eid}{016125} (pages~\bibinfo{numpages}{8}) (\bibinfo{year}{2006}).

\bibitem[{\citenamefont{Holme et~al.}(2002)\citenamefont{Holme, Kim, Yoon, and
  Han}}]{holme}
\bibinfo{author}{\bibfnamefont{P.}~\bibnamefont{Holme}},
  \bibinfo{author}{\bibfnamefont{B.~J.} \bibnamefont{Kim}},
  \bibinfo{author}{\bibfnamefont{C.~N.} \bibnamefont{Yoon}}, \bibnamefont{and}
  \bibinfo{author}{\bibfnamefont{S.~K.} \bibnamefont{Han}},
  \bibinfo{journal}{Physical Review E} \textbf{\bibinfo{volume}{65}},
  \bibinfo{pages}{056109} (\bibinfo{year}{2002}).

\bibitem[{\citenamefont{R.~Albert and Barab��si}(2000)}]{albert}
\bibinfo{author}{\bibfnamefont{H.~J.} \bibnamefont{R.~Albert}}
  \bibnamefont{and} \bibinfo{author}{\bibfnamefont{A.-L.}
  \bibnamefont{Barab��si}}, \bibinfo{journal}{Nature (London)}
  \textbf{\bibinfo{volume}{406}}, \bibinfo{pages}{198701}
  (\bibinfo{year}{2000}).

\bibitem[{\citenamefont{Karrer et~al.}(2008{\natexlab{a}})\citenamefont{Karrer,
  Levina, and Newman}}]{karrer}
\bibinfo{author}{\bibfnamefont{B.}~\bibnamefont{Karrer}},
  \bibinfo{author}{\bibfnamefont{E.}~\bibnamefont{Levina}}, \bibnamefont{and}
  \bibinfo{author}{\bibfnamefont{M.~E.~J.} \bibnamefont{Newman}},
  \bibinfo{journal}{Physical Review E} \textbf{\bibinfo{volume}{77}},
  \bibinfo{eid}{046119} (pages~\bibinfo{numpages}{9})
  (\bibinfo{year}{2008}{\natexlab{a}}).

\bibitem[{\citenamefont{Moreira et~al.}(2009)\citenamefont{Moreira, Jos\'{e}
  S.~Andrade, Herrmann, and Indekeu}}]{moreira}
\bibinfo{author}{\bibfnamefont{A.~A.} \bibnamefont{Moreira}},
  \bibinfo{author}{\bibfnamefont{J.}~\bibnamefont{Jos\'{e} S.~Andrade}},
  \bibinfo{author}{\bibfnamefont{H.~J.} \bibnamefont{Herrmann}},
  \bibnamefont{and} \bibinfo{author}{\bibfnamefont{J.~O.}
  \bibnamefont{Indekeu}}, \bibinfo{journal}{Physical Review Letters}
  \textbf{\bibinfo{volume}{102}}, \bibinfo{pages}{018701}
  (\bibinfo{year}{2009}).

\bibitem[{\citenamefont{Paolo~Crucitti and Rapisarda}(2004)}]{paolo}
\bibinfo{author}{\bibfnamefont{M.~M.} \bibnamefont{Paolo~Crucitti},
  \bibfnamefont{Vito~Latorab}} \bibnamefont{and}
  \bibinfo{author}{\bibfnamefont{A.}~\bibnamefont{Rapisarda}},
  \bibinfo{journal}{Physica A} \textbf{\bibinfo{volume}{430}},
  \bibinfo{eid}{018701} (\bibinfo{year}{2004}).

\bibitem[{\citenamefont{Herrmann et~al.}(2010)\citenamefont{Herrmann,
  Schneider, Moreira, Jr, and Havlin}}]{onion}
\bibinfo{author}{\bibfnamefont{H.~J.} \bibnamefont{Herrmann}},
  \bibinfo{author}{\bibfnamefont{C.~M.} \bibnamefont{Schneider}},
  \bibinfo{author}{\bibfnamefont{A.~A.} \bibnamefont{Moreira}},
  \bibinfo{author}{\bibfnamefont{J.~S.~A.} \bibnamefont{Jr}}, \bibnamefont{and}
  \bibinfo{author}{\bibfnamefont{S.}~\bibnamefont{Havlin}},
  \bibinfo{journal}{Journal of Statistical Mechanics: Theory and Experiment (In
  Press)}  (\bibinfo{year}{2010}).

\bibitem[{\citenamefont{Buldyrev et~al.}(2010)\citenamefont{Buldyrev, Parshani,
  Paul, Stanley, and Havlin}}]{cata}
\bibinfo{author}{\bibfnamefont{S.~V.} \bibnamefont{Buldyrev}},
  \bibinfo{author}{\bibfnamefont{R.}~\bibnamefont{Parshani}},
  \bibinfo{author}{\bibfnamefont{G.}~\bibnamefont{Paul}},
  \bibinfo{author}{\bibfnamefont{H.~E.} \bibnamefont{Stanley}},
  \bibnamefont{and} \bibinfo{author}{\bibfnamefont{S.}~\bibnamefont{Havlin}},
  \bibinfo{journal}{Nature} \textbf{\bibinfo{volume}{464}},
  \bibinfo{pages}{1025} (\bibinfo{year}{2010}), ISSN \bibinfo{issn}{0028-0836}.

\bibitem[{\citenamefont{Xiao et~al.}(2008)\citenamefont{Xiao, Xiao, and
  Cheng}}]{comm}
\bibinfo{author}{\bibfnamefont{S.}~\bibnamefont{Xiao}},
  \bibinfo{author}{\bibfnamefont{G.}~\bibnamefont{Xiao}}, \bibnamefont{and}
  \bibinfo{author}{\bibfnamefont{T.~H.} \bibnamefont{Cheng}},
  \bibinfo{journal}{Communications Magazine, IEEE}
  \textbf{\bibinfo{volume}{46}}, \bibinfo{pages}{146 } (\bibinfo{year}{2008}).

\bibitem[{\citenamefont{Wu et~al.}(2007)\citenamefont{Wu, Deng, Tan, and
  Zhu}}]{incomp}
\bibinfo{author}{\bibfnamefont{J.}~\bibnamefont{Wu}},
  \bibinfo{author}{\bibfnamefont{H.~Z.} \bibnamefont{Deng}},
  \bibinfo{author}{\bibfnamefont{Y.~J.} \bibnamefont{Tan}}, \bibnamefont{and}
  \bibinfo{author}{\bibfnamefont{D.~Z.} \bibnamefont{Zhu}},
  \bibinfo{journal}{Journal of Physics A: Mathematical and Theoretical}
  \textbf{\bibinfo{volume}{40}}, \bibinfo{pages}{2665} (\bibinfo{year}{2007}).

\bibitem[{\citenamefont{Karrer et~al.}(2008{\natexlab{b}})\citenamefont{Karrer,
  Levina, and Newman}}]{rob_comm}
\bibinfo{author}{\bibfnamefont{B.}~\bibnamefont{Karrer}},
  \bibinfo{author}{\bibfnamefont{E.}~\bibnamefont{Levina}}, \bibnamefont{and}
  \bibinfo{author}{\bibfnamefont{M.~E.~J.} \bibnamefont{Newman}},
  \bibinfo{journal}{Physical Review E} \textbf{\bibinfo{volume}{77}},
  \bibinfo{pages}{046119} (\bibinfo{year}{2008}{\natexlab{b}}).

\bibitem[{\citenamefont{Newman and Girvan}(2004)}]{find_comm}
\bibinfo{author}{\bibfnamefont{M.~E.~J.} \bibnamefont{Newman}}
  \bibnamefont{and} \bibinfo{author}{\bibfnamefont{M.}~\bibnamefont{Girvan}},
  \bibinfo{journal}{Physical Review E} \textbf{\bibinfo{volume}{69}},
  \bibinfo{pages}{026113} (\bibinfo{year}{2004}).

\bibitem[{\citenamefont{Newman}(2004)}]{alg_comm}
\bibinfo{author}{\bibfnamefont{M.~E.~J.} \bibnamefont{Newman}},
  \bibinfo{journal}{Physical Review E} \textbf{\bibinfo{volume}{69}},
  \bibinfo{pages}{066133} (\bibinfo{year}{2004}).

\bibitem[{\citenamefont{Lemke et~al.}(2004)\citenamefont{Lemke, Her\'{e}dia,
  Barcellos, Dos~Reis, and Mombach}}]{dam-meta}
\bibinfo{author}{\bibfnamefont{N.}~\bibnamefont{Lemke}},
  \bibinfo{author}{\bibfnamefont{F.}~\bibnamefont{Her\'{e}dia}},
  \bibinfo{author}{\bibfnamefont{C.~K.} \bibnamefont{Barcellos}},
  \bibinfo{author}{\bibfnamefont{A.~N.} \bibnamefont{Dos~Reis}},
  \bibnamefont{and} \bibinfo{author}{\bibfnamefont{J.~C.~M.}
  \bibnamefont{Mombach}}, \bibinfo{journal}{Bioinformatics}
  \textbf{\bibinfo{volume}{20}}, \bibinfo{pages}{115} (\bibinfo{year}{2004}),
  ISSN \bibinfo{issn}{1367-4803}.

\bibitem[{\citenamefont{Schmith et~al.}(2005)\citenamefont{Schmith, Lemke,
  Mombach, Benelli, Barcellos, and Bedin}}]{dam-ppi}
\bibinfo{author}{\bibfnamefont{J.}~\bibnamefont{Schmith}},
  \bibinfo{author}{\bibfnamefont{N.}~\bibnamefont{Lemke}},
  \bibinfo{author}{\bibfnamefont{J.~C.} \bibnamefont{Mombach}},
  \bibinfo{author}{\bibfnamefont{P.}~\bibnamefont{Benelli}},
  \bibinfo{author}{\bibfnamefont{C.~K.} \bibnamefont{Barcellos}},
  \bibnamefont{and} \bibinfo{author}{\bibfnamefont{G.~B.} \bibnamefont{Bedin}},
  \bibinfo{journal}{Physica A: Statistical Mechanics and its Applications}
  \textbf{\bibinfo{volume}{349}}, \bibinfo{pages}{675 } (\bibinfo{year}{2005}),
  ISSN \bibinfo{issn}{0378-4371}.

\bibitem[{\citenamefont{Barabasi and Albert}(1999)}]{ba}
\bibinfo{author}{\bibfnamefont{A.~L.} \bibnamefont{Barabasi}} \bibnamefont{and}
  \bibinfo{author}{\bibfnamefont{R.}~\bibnamefont{Albert}},
  \bibinfo{journal}{Science (New York, N.Y.)} \textbf{\bibinfo{volume}{286}},
  \bibinfo{pages}{509} (\bibinfo{year}{1999}), ISSN \bibinfo{issn}{1095-9203}.

\bibitem[{\citenamefont{Erd{\H{o}}s and R{\'{e}}nyi}(1959)}]{er}
\bibinfo{author}{\bibfnamefont{P.}~\bibnamefont{Erd{\H{o}}s}} \bibnamefont{and}
  \bibinfo{author}{\bibfnamefont{A.}~\bibnamefont{R{\'{e}}nyi}},
  \bibinfo{journal}{Publ. Math. Debrecen} \textbf{\bibinfo{volume}{6}},
  \bibinfo{pages}{290} (\bibinfo{year}{1959}).

\bibitem[{\citenamefont{Holme}(2004)}]{holmeepl}
\bibinfo{author}{\bibfnamefont{P.}~\bibnamefont{Holme}}, \bibinfo{journal}{EPL
  (Europhysics Letters)} \textbf{\bibinfo{volume}{68}}, \bibinfo{pages}{908}
  (\bibinfo{year}{2004}),
  \urlprefix\url{http://stacks.iop.org/0295-5075/68/i=6/a=908}.

\bibitem[{\citenamefont{Holme et~al.}(2007)\citenamefont{Holme, Karlin, and
  Forrest}}]{holme07}
\bibinfo{author}{\bibfnamefont{P.}~\bibnamefont{Holme}},
  \bibinfo{author}{\bibfnamefont{J.}~\bibnamefont{Karlin}}, \bibnamefont{and}
  \bibinfo{author}{\bibfnamefont{S.}~\bibnamefont{Forrest}}, in
  \emph{\bibinfo{booktitle}{Proceedings of the Royal Society A 463}}
  (\bibinfo{year}{2007}), pp. \bibinfo{pages}{1231--1246}.

\bibitem[{\citenamefont{Brandes}(2001)}]{fastbw}
\bibinfo{author}{\bibfnamefont{U.}~\bibnamefont{Brandes}},
  \bibinfo{journal}{Journal of Mathematical Sociology}
  \textbf{\bibinfo{volume}{25}}, \bibinfo{pages}{163} (\bibinfo{year}{2001}).

\bibitem[{\citenamefont{Whitney}(1932)}]{whitney}
\bibinfo{author}{\bibfnamefont{H.}~\bibnamefont{Whitney}},
  \bibinfo{journal}{Transactions of the American Mathematical Society}
  \textbf{\bibinfo{volume}{34}}, \bibinfo{pages}{339} (\bibinfo{year}{1932}).

\bibitem[{\citenamefont{Gross and Yellen}(2005)}]{graph}
\bibinfo{author}{\bibfnamefont{J.~L.} \bibnamefont{Gross}} \bibnamefont{and}
  \bibinfo{author}{\bibfnamefont{J.}~\bibnamefont{Yellen}},
  \emph{\bibinfo{title}{Graph Theory and Its Applications, Second Edition
  (Discrete Mathematics and Its Applications)}} (\bibinfo{publisher}{Chapman \&
  Hall/CRC}, \bibinfo{year}{2005}), ISBN \bibinfo{isbn}{158488505X}.

\end{thebibliography}

\end{document}